\documentclass[
 aip,
 pop,
 amsmath,
 amssymb,
 reprint,
 longbibliography,
 floatfix
]{revtex4-2}

\usepackage{amsmath}
\usepackage{amssymb}
\usepackage{graphicx}
\usepackage{bm}
\usepackage{dcolumn}
\usepackage{color}
\usepackage{placeins}
\usepackage{array}
\usepackage{hyperref}

\graphicspath{{figures/}}

\newcommand{\singlefigwidth}{\linewidth}
\newcommand{\widefigwidth}{0.94\textwidth}
\newcommand{\tablecaptiongap}{\vspace{0.45em}}
\newcolumntype{L}[1]{>{\raggedright\arraybackslash}p{#1}}


\begin{document}

\title{
Revisiting the MRX Electron Current Sheet Width with Semi-Collisional Kinetic Simulations at Hydrogen Mass Ratio
}

\author{Sung Hyun Son}
\email{sson2@pppl.gov}
\affiliation{\mbox{Department of Astrophysical Sciences, Princeton University, Princeton, New Jersey, USA}}
\affiliation{Princeton Plasma Physics Laboratory, Princeton, New Jersey, USA}

\author{Adam Stanier}
\affiliation{Los Alamos National Laboratory, Los Alamos, New Mexico, USA}

\author{William Daughton}
\affiliation{Los Alamos National Laboratory, Los Alamos, New Mexico, USA}

\author{Jongsoo Yoo}
\affiliation{Princeton Plasma Physics Laboratory, Princeton, New Jersey, USA}

\author{Tongnyeol Rhee}
\affiliation{Korea Institute of Fusion Energy, Daejeon, Republic of Korea}

\author{Hantao Ji}
\affiliation{\mbox{Department of Astrophysical Sciences, Princeton University, Princeton, New Jersey, USA}}
\affiliation{Princeton Plasma Physics Laboratory, Princeton, New Jersey, USA}
\date{\today}

\makeatletter
\ifdefined\revtexpreview@maketitle
  \author{Sung Hyun Son$^{1,2}$, Adam Stanier$^{3}$, William Daughton$^{3}$,\\
  Jongsoo Yoo$^{1}$, Tongnyeol Rhee$^{4}$, and Hantao Ji$^{1,2}$\\[0.35em]
  {\small $^{1}$Princeton Plasma Physics Laboratory, Princeton, New Jersey 08543, USA}\\
  {\small $^{2}$Department of Astrophysical Sciences, Princeton University, Princeton, New Jersey 08544, USA}\\
  {\small $^{3}$Los Alamos National Laboratory, Los Alamos, New Mexico 87545, USA}\\
  {\small $^{4}$Korea Institute of Fusion Energy, Daejeon 34133, Republic of Korea}}%
\fi
\makeatother

\begin{abstract}
For over eighteen years, the electron current sheet measured in the Magnetic Reconnection Experiment (MRX) has stood a factor of 2--5 wider than predicted by MRX-like kinetic simulations, and the measured electron force balance has not closed with the classical terms alone. As a consequence, the dominant nonideal terms responsible for breaking the frozen-in condition have remained unexplained. Here, two-dimensional kinetic simulations with binary Coulomb collisions are performed in a cylindrical geometry representative of MRX, at the realistic hydrogen mass ratio $m_i/m_e = 1836$ and at MRX-relevant collisionality. For the first time, these simulations reproduce the measured electron current sheet half-width. The simulated value, $\delta_{BT} = 0.744 \pm 0.054$~cm or $6.26 \pm 0.45$ electron skin depths ($d_e$), lies within the experimental range of 5.5--7.5~$d_e$. The electron force balance closes through the classical channels alone: the pressure-tensor divergence supports 76\% of the nonideal electric field and collisional friction the remainder. The historical force-balance deficit reappears only when the simulated layer is sampled at the experimental 3~cm outflow resolution, suggesting that the deficit reflects probe resolution rather than anomalous dissipation. Beyond this reproduction, an analytic model of the layer width is developed that orders the meandering electrons by the coherence of their orbits against collisions. In this model, the Dreicer ratio $E_D/|E_y|$ selects the electrons whose current-carrying motion survives, and the resulting width prediction brackets the measured values across a wide collisionality scan. A discrepancy remains in the width normalized to the local electron gyroradius ($\rho_e$), whose measured value lies a factor of 3--6 above both the model and the simulations. A controlled simulation with an additional electron heat sink traces part of this gap to the local thermodynamic state of the current sheet, leaving the unmodeled transport and loss channels, as well as finer experimental electron-layer measurements, for future work.
\end{abstract}

\maketitle

\section{Introduction}
\label{sec:intro}

The electron diffusion region (EDR) is the kinetic layer where electrons decouple from the reconnecting magnetic field, breaking the frozen-in constraint and enabling magnetic topology change. While the global reconnection rate is governed by ion-scale dynamics, the electron layer adjusts to balance the nonideal electric field responsible for this topological breaking. Understanding the detailed plasma physics of this layer has been a major science goal, notably driving the Magnetospheric Multiscale (MMS) mission \cite{Burch2016SSR}. Near the X line, this nonideal electric field can be supported by the divergence of the electron pressure tensor, $\nabla\cdot{\bf P}_e$, specifically through its off-diagonal or nongyrotropic components \cite{Hesse1999PoP,Kuznetsova2001JGR}. At the particle level, meandering electron orbits in the reversing magnetic field \cite{Speiser1965JGR,Sonnerup1971JGR,Buchner1989JGR} physically provide this tensor support. Consequently, a long lineage of kinetic simulations, spanning from driven \cite{Horiuchi1994PoP} to spontaneous Harris-sheet reconnection \cite{Le2013PRL,Egedal2023PoP,Egedal2024GRL}, has directly linked the structure and width of the electron current sheet to the meandering-orbit scale.

However, when applied to the Magnetic Reconnection Experiment (MRX) \cite{Yamada1997MRX} operating in the semi-collisional regime, this standard paradigm has faced two long-standing discrepancies. First, the measured electron-layer half-width (probe-corrected to $5.5$--$7.5$ electron skin depths, $d_e$) exceeded the $1.5$--$3\,d_e$ widths predicted by two-dimensional kinetic simulations at reduced mass ratio by a factor of two to five \cite{Ren2008PRL,Ji2008GRL,Dorfman2008PoP,Roytershteyn2010PoP}. Second, the measured reconnection electric field greatly exceeded the sum of the classical electron-layer dissipative channels, namely the nongyrotropic pressure estimate of Hesse et al.\ \cite{Hesse1999PoP} and Spitzer friction, leaving the majority of the nonideal field unaccounted for \cite{Ji2008GRL}. Both discrepancies were widely suspected to arise from physics beyond the two-dimensional (axisymmetric) terms in Ohm's law. Dorfman et al.~\cite{Dorfman2008PoP} identified the electron-layer discrepancy in collisionless simulations that already reproduced the ion-scale features of MRX, and named weak Coulomb collisions and three-dimensional effects as the two candidates, the latter through the lower-hybrid electromagnetic fluctuations long measured in the device \cite{Ji2004PRL,Carter2001PRL,Daughton2003PoP}. Both have since been tested, though only at reduced mass ratio, which makes the collisionality of the experiment difficult to scale properly: collisions widen the layer, but not enough to close the gap at the relevant collisionality \cite{Roytershteyn2010PoP}, and lower-hybrid fluctuations at or above the measured amplitudes broadened the toroidally averaged layer by only about 25\% \cite{Roytershteyn2013PoP}. All of these simulations also used Cartesian domains that carry the MRX boundary conditions without the toroidal geometry of the device \cite{Dorfman2008PoP,Roytershteyn2010PoP,Roytershteyn2013PoP}, so the collisional case is worth revisiting once these restrictions are lifted.

To resolve these lingering discrepancies, it is essential to revisit the fundamental physics of driven magnetic reconnection in the two-dimensional, semi-collisional antiparallel regime. The natural governing parameter here is the ratio of the Dreicer field to the reconnection electric field, $E_D/|E_y|$, which quantifies how effectively Coulomb collisions regulate the formation of non-thermal electron tails (Braginskii-type transport requires $|E_y|\ll E_D$) \cite{Dreicer1959PR,Roytershteyn2010PoP}. While collisional particle-in-cell studies have shown a transition from collisional to kinetic reconnection \cite{Daughton2009PRL} that is organized by a mass-ratio-scaled form of this same ratio \cite{JaraAlmonte2021PRL}, a comprehensive structural model remains elusive for antiparallel geometries. For strong guide-field reconnection, reduced fluid-kinetic theory successfully provides analytic tearing-layer widths across all collisionality regimes \cite{Zocco2011PoP,Loureiro2013PRL}, revealing a two-scale current layer \cite{Stanier2015PoP}. However, no equivalent analytic ordering exists for the semi-collisional antiparallel case, where neither the collisionless meandering-orbit width nor the resistive Sweet--Parker thickness directly applies.

In this paper, we address these gaps by revisiting both long-standing discrepancies using two-dimensional VPIC simulations. These simulations include binary Coulomb collisions, model the flux-core drive and boundaries of the MRX cylindrical geometry, and, for the first time, achieve the realistic proton-to-electron mass ratio at MRX-relevant collisionality. Our comparison with experimental data focuses on the local layer width and the electron force balance. Furthermore, we pair these macroscopic comparisons with a microscopic orbit-level ordering model. This model builds upon the Maxwellian-averaged electron meandering width of Matsui and Daughton \cite{Matsui2008PoP}, expanding it to account for finite collisional effects.

The paper is organized as follows. Section~\ref{sec:dataset} describes the VPIC simulation setup and the width diagnostics. Section~\ref{sec:mrx_benchmark} presents the reproduction of the MRX layer: the local width, the electron force balance, and the force-balance accounting. Section~\ref{sec:model} traces these results to the electron orbits, identifying the current-carrying population, deriving a coherence ordering for meandering electrons under collisions, and testing the resulting analytic width model against a collisionality scan, which also shows the width and the force-balance partition are organized by different variables. Section~\ref{sec:discussion} reports the remaining $\rho_e$-normalized discrepancy and examines its relation to the local thermodynamic state of the current sheet with a simulation with an imposed electron cooling. Section~\ref{sec:conclusions} summarizes the conclusions.

\section{VPIC Simulations and Diagnostics}
\label{sec:dataset}

\subsection{Simulation setup}
\label{sec:setup}

\begin{figure*}[!t]
\centering
\includegraphics[width=\widefigwidth]{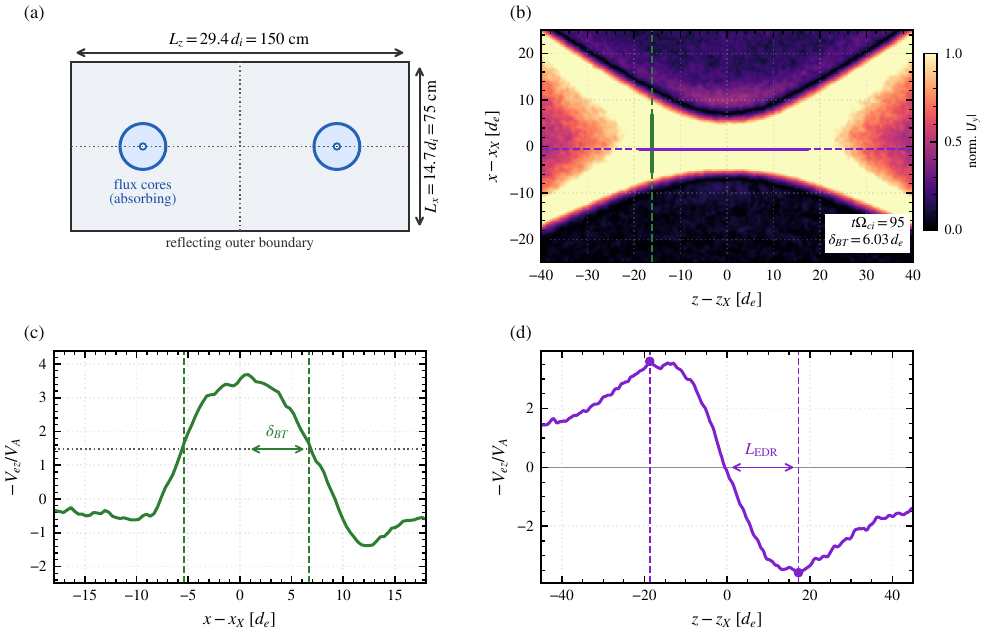}
\caption{Cylindrical simulation domain modeled on MRX and local electron-layer diagnostics in the $m_i/m_e=1836$ collisional VPIC run. (a) Domain and flux-core geometry. (b) $|J_y|$, normalized to its in-plane maximum, near the X line, with the cross-sheet cut (green, taken at the plane of peak electron outflow) and the outflow cut (purple, along the current center); the inset quotes $\delta_{BT}$ for this single snapshot, whereas Table~\ref{tab:metric_comparison} quotes the mean over $t\Omega_{ci}=80$--110. The vertical coordinate is measured from the X-line position $x_X$ of Eq.~\eqref{eq:harris_fit}. (c) Cross-sheet $-V_{ez}(x)/V_A$ cut defining the 40\% electron-outflow half-width $\delta_{BT}$, plotted against $x-x_X$. (d) Outflow $-V_{ez}(z)/V_A$ cut defining the EDR half-length $L_{\rm EDR}$. The normalization $V_A$ is the Alfv\'en speed evaluated with the Harris-shoulder field $B_H^*$ of Eq.~\eqref{eq:harris_fit} and the local density at the sheet center. Panels (b)--(d) show a representative output at $t\Omega_{ci}=95$.}
\label{fig:local_diagnostics}
\end{figure*}

All simulations in this paper use VPIC, an explicit electromagnetic particle-in-cell code that advances electron and ion macroparticles and deposits a charge-conserving current on a finite-difference time-domain field mesh \cite{Bowers2008PoP,VPIC}. The primary run has $m_i/m_e=1836$, binary Coulomb collisions, and a cylindrical domain whose flux-core drive and boundary conditions are modeled on MRX (Fig.~\ref{fig:local_diagnostics}(a)). The simulation is two dimensional in the $(x,z)$ plane. Here $x$ is the inflow and cross-sheet direction (the experimental radial direction $R$), $z$ the outflow direction (the axial $Z$), and $y$ the out-of-plane toroidal direction ($\theta$) of the current and the reconnection electric field. We retain the $(x,y,z)$ labels for continuity with prior MRX-oriented kinetic simulations \cite{Dorfman2008PoP,Roytershteyn2010PoP}. Cylindrical geometry means that the field solve and particle push are performed in the $(R,Z)$ plane of the experiment with the corresponding metric factors, and that the flux-core and outer boundaries sit at their MRX positions, rather than in a Cartesian box. This cylindrical version of VPIC has previously been validated against TREX experimental data \cite{Greess2021JGR,Greess2022PoP}. The primary-run domain spans $L_x\times L_z=14.7\times29.4\,d_i$, with $d_i$ the ion inertial length at the reference density ($631\times1262\,d_e$ or $75\times150$ cm), and the $d_i$-normalized box size is held approximately constant across the mass-ratio scan so that the ion-scale reconnection geometry is comparable across runs. The outer $x$ and $z$ domain boundaries use conducting field boundaries and reflecting particle boundaries, and the flux-core particle regions are absorbing. Because these runs are the first published application of VPIC in a cylindrical MRX configuration, Sec.~\ref{sec:cyl_control} compares matched Cartesian and cylindrical runs. For semi-collisional simulation, the Takizuka--Abe binary Coulomb operator \cite{Takizuka1977JCP} is applied every five VPIC time steps to electron--electron, electron--ion, and ion--ion pairs. 

We model only the pull phase of MRX-type driven reconnection. The coil current density ramps down as $j(t)=j_0\left[1+5\cos^2(\pi t/2\tau)\right]/6$, the waveform of Refs.~\cite{Dorfman2008PoP,Roytershteyn2010PoP}, with the ramp timescale $\tau=150\Omega_{ci}^{-1}$ held fixed across all runs. Here, $\Omega_{ci}$ is the ion cyclotron frequency at the reference field $B_0$, defined as the magnetic field $3\,d_i$ upstream of the X line along the $x$ inflow line, to keep the global drive comparable, following the approach of prior MRX-oriented kinetic simulations \cite{Dorfman2008PoP,Roytershteyn2010PoP}. We define the quasi-steady diagnostic interval as $t\Omega_{ci}=80$--110, over which the reconnection rate varies by less than 10\%, the current sheet structure is steady, and no secondary islands form. Unless otherwise stated, VPIC uncertainties are standard deviations across this interval.

The physical-unit mapping uses the reference fill density $n_0=2\times10^{13}~{\rm cm}^{-3}$ and $T_e=T_i=5$ eV, characteristic of the hydrogen MRX discharges used for comparison. These references give a thermal electron--ion collision rate $\nu_{ei}^{\rm th}$ with $\nu_{ei}^{\rm th}/\Omega_{ce}=0.0197$, where $\Omega_{ce}$ is the electron cyclotron frequency at the reference field $B_0$; an electron inertial length $d_e=0.119$ cm; and an initial upstream $\beta_{e,0}\simeq0.075$ at the reference point where $B_0$ is obtained, where the subscript 0 denotes the initial reference values. The upstream value at the same location rises to $\simeq0.73$ in quasi-steady state. Throughout, MRX-relevant collisionality refers to this imposed thermal rate, defined with the initial reference values, rather than to the time-dependent Dreicer ratio that the layer develops dynamically. Dimensional quantities are reported in Gaussian-cgs units unless explicitly stated. Following the approach of prior MRX kinetic simulations, which reduce $\omega_{pe}/\Omega_{ce}$ from its MRX value of $\simeq70$ to order unity at fixed $\beta_e$ \cite{Dorfman2008PoP}, we set $\omega_{pe}/\Omega_{ce}=1$, with $\omega_{pe}$ the electron plasma frequency at the reference density $n_0$.

Table~\ref{tab:run_parameters} summarizes the simulation runs used in this paper, all evaluated with the same width diagnostics of Sec.~\ref{sec:diagnostics} and the same coherence-ordering procedure of Sec.~\ref{sec:model}. The primary run is the realistic-mass-ratio collisional case used for the experimental benchmark. The collisional scan tests whether the same semi-collisional width ordering organizes cases with varied mass ratio, collisionality, and local thermodynamic state, with the per-case parameters available as tabulated data (see Data Availability). Individual control runs used for single comparisons, such as the collisionless companion of Sec.~\ref{sec:vspace_complete}, are described where they appear. The $m_i/m_e=100$ cooling-control sequence imposes an ad hoc electron energy sink whose strength is scanned; its form and purpose are described in Sec.~\ref{sec:cooling_control}, where the sequence is used to test how the layer responds as its thermodynamic state changes.

\begin{table*}[!htb]
\centering
\caption{Summary of simulation runs. All runs share the cylindrical MRX-modeled configuration, cell size $0.37\,d_e=2\lambda_{De}$, with $\lambda_{De}$ the electron Debye length, time step $\omega_{pe}\Delta t=0.129$ (half the cylindrical Courant limit). The ppc column lists the average number of macroparticles per cell per species.}
\label{tab:run_parameters}
\tablecaptiongap
\small
\setlength{\tabcolsep}{3.0pt}
\renewcommand{\arraystretch}{1.08}
\begin{tabular}{L{0.16\textwidth}L{0.09\textwidth}L{0.34\textwidth}L{0.16\textwidth}L{0.05\textwidth}}
\hline
Run & $m_i/m_e$ & Reference $\nu_{ei}^{\rm th}/\Omega_{ce}$ \& control term & Grid & ppc \\
\hline
Primary & 1836 & $\nu_{ei}^{\rm th}/\Omega_{ce}=0.0197$ & $1728\times1\times3472$ & 180 \\
Collisional scan & 100--1836 & $\nu_{ei}^{\rm th}/\Omega_{ce}=0.003$--0.24 & varies by case & 180 \\
Cooling control & 100 & $\nu_{ei}^{\rm th}/\Omega_{ce}=0.0394$, + electron $T_e^4$ cooling & $400\times1\times784$ & 200 \\
\hline
\end{tabular}
\end{table*}

\subsection{Width and normalization diagnostics}
\label{sec:diagnostics}

Figure~\ref{fig:local_diagnostics} shows the primary run geometry and the local diagnostic cuts used throughout the paper. The $|J_y|$ map identifies the central current layer, the $V_{ez}(x)$ cut defines the 40\% electron-outflow half-width $\delta_{BT}$, and the $V_{ez}(z)$ cut defines $L_{\rm EDR}$. The scan runs use the same width definitions. With the coordinate convention defined above, all cross-sheet widths are half-widths measured along $x$. We use thickness and width interchangeably, always quoting half-widths.

The direct MRX comparison uses the $40\%$ electron-outflow half-width $\delta_{BT}$, with the subscript inherited from the notation of Refs.~\cite{Dorfman2008PoP,RenThesis}. We evaluate this diagnostic on a cross-sheet cut through the downstream location where the electron outflow speed $|V_{ez}|$ is largest, and define $\delta_{BT}$ as the half-width at which $|V_{ez}|$ falls to 40\% of its peak value on that cut. This is the same definition used to measure the electron-layer thickness in the MRX experimental data \cite{Dorfman2008PoP,RenThesis}. The EDR half-length $L_{\rm EDR}$ is half the $z$ separation between the positive and negative $V_{ez}$ extrema on the outflow cut. We also measure the central current layer half-width $\delta_c$, obtained at the X-line reconnecting-field reversal from a Harris-plus-linear fit \cite{Roytershteyn2010PoP},
\begin{equation}
  B_z(x)=B_H^*\tanh\left(\frac{x-x_X}{\delta_c}\right)
  +C(x-x_X)+B_{\rm bg}.
  \label{eq:harris_fit}
\end{equation}
Here $x_X$ is the fitted X-line, or Harris-reversal, center. The linear term $C(x-x_X)+B_{\rm bg}$ accounts for the background-field contribution, characteristic of driven reconnection experiments. The two widths differ both by definition and by location. $\delta_{BT}$ measures the downstream electron-outflow layer, while $\delta_c$ measures the local magnetic-gradient/central-layer scale at the X-line reversal. For the rectangular-like layers in this work, Fig.~\ref{fig:sheetwidth} of Appendix~\ref{app:fit_width} shows that $\delta_{BT}$ and $\delta_c$ follow the same trend across the scan, differing by about 20\% for the primary run.

The normalization scales are
\begin{equation}
  d_e=\frac{c}{\omega_{pe}(n_0)}, \qquad
  \rho_e=\frac{v_{the}}{\Omega_{ce}(B_H^*)},
\end{equation}
where $d_e$ is evaluated at the run-reference density $n_0$ (0.119 cm for the primary run) and $\rho_e$ is evaluated using the local Harris-shoulder field $B_H^*$ and the local electron thermal speed $v_{the}$. In Gaussian-cgs units, $\omega_{pe}(n_e)=(4\pi n_e e^2/m_e)^{1/2}$, $\Omega_{ce}(B)=eB/(m_ec)$, and the electron thermal speed is
\begin{equation}
  v_{the}=\left(\frac{p_e}{n_em_e}\right)^{1/2}
  =\left(\frac{T_e}{m_e}\right)^{1/2},
\end{equation}
with $p_e$ the local scalar electron pressure, $p_e={\rm tr}({\bf P}_e)/3$, at the X-line current center. Because the magnetic field vanishes at the X line, the fitted amplitude $B_H^*$ of Eq.~\eqref{eq:harris_fit} serves as the magnetic field reference for $\rho_e$ and for the local scalar beta,
\begin{equation}
  \beta_e=\frac{8\pi p_e}{(B_H^*)^2}.
\end{equation}
With these conventions the exact relation is
\begin{equation}
  \frac{\rho_e}{d_e}
  =\left(\frac{\beta_e}{2}\right)^{1/2}
   \left(\frac{n_0}{n_e}\right)^{1/2},
  \label{eq:rhoe_beta_relation}
\end{equation}
where $n_e$ is the local X-line density. For the primary run in quasi-steady state, the X line is depleted to $n_e\simeq0.7\,n_0$, with $\beta_e\simeq3.2$ and Eq.~\eqref{eq:rhoe_beta_relation} giving $\rho_e/d_e\simeq1.51$. In the simulations presented in this paper, the current sheets form dynamically in response to the coil drive, and differences in temperature, density, field profile, or distribution-function-sensitive thermal speed can make a dimensionally similar layer appear narrower or wider in $\rho_e$ units.

The Dreicer ratio $E_D/|E_y|$ enters the coherence ordering of Sec.~\ref{sec:model}. We evaluate $E_D$ from the run-controlled inputs as
\begin{equation}
  E_D=\frac{m_e\,\nu_{ei}^{\rm th}\,v_{the,0}}{e},
  \label{eq:ed_operational}
\end{equation}
where $\nu_{ei}^{\rm th}$ is the thermal electron--ion collision rate and $v_{the,0}=\sqrt{T_{e,0}/m_e}$ is the initial electron thermal speed, both taken at the initial reference density and temperature ($\nu_{ei}^{\rm th}/\Omega_{ce}=0.0197$ for the primary run, the ratio quoted in Table~\ref{tab:run_parameters}). This is the conventional Dreicer field \cite{Dreicer1959PR}, evaluated at the initial reference inputs that set the run-controlled collisionality. The field $E_y$ is the out-of-plane electric field averaged in time over $\approx500\,\Omega_{ce}^{-1}$ and in space over a box of one-$d_e$ half-width in $x$ and $z$ centered on the X-line current center. Throughout, $\chi\equiv E_D/|E_y|$ denotes the Dreicer ratio.

\subsection{Impact of cylindrical geometry}
\label{sec:cyl_control}

\begin{figure*}[!t]
\centering
\includegraphics[width=\widefigwidth]{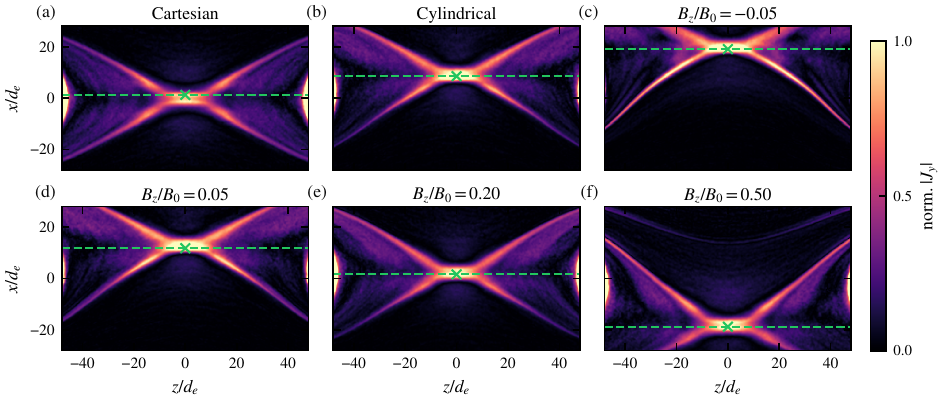}
\caption{Stacked out-of-plane current maps for reduced-mass controls at $m_i/m_e=150$. Each panel shows the current density $J_y$ in the sense of the main sheet current, normalized to its in-plane maximum, at $t\Omega_{ci}=95$. (a) Cartesian control. (b) Cylindrical control (with zero additional $B_z$ equilibrium field imposed). (c)--(f) Selected imposed-$B_z$ cylindrical cases from the EF control, with the imposed values quoted in units of the reference field $B_0$. The green dashed line and cross mark the peak-current position in each panel for reference.}
\label{fig:sheet_motion}
\end{figure*}

To isolate directly what the coordinate change does and does not alter in the layer diagnostics, the geometry control compares collisionless Cartesian and cylindrical runs at a single reduced-mass case, $m_i/m_e=150$. The local widths remain mostly unchanged, $\delta_{BT}=2.44\pm0.14$ and $2.55\pm0.15\,d_e$, respectively. The X-line center and the current sheet are radially displaced in the cylindrical run, as shown in Figs.~\ref{fig:sheet_motion}(a) and \ref{fig:sheet_motion}(b). The displacement is the expected cylindrical effect. Now the current sheet is a toroidal current ring, and the poloidal flux it generates is compressed on its inboard side, where the volume is smaller. The magnetic pressure is therefore higher inboard than outboard, and the sheet feels a net outward (hoop) force that has no counterpart in the mirror-symmetric Cartesian geometry. In MRX the same tendency is compensated by the uniform axial field of the Equilibrium Field (EF) coils, whose $\mathbf{J}\times\mathbf{B}$ force on the toroidal current pushes the sheet back inboard.

The EF control reproduces this compensation by imposing a uniform $B_z$ offset [Figs.~\ref{fig:sheet_motion}(c)--(f)]: as the offset increases from $-0.05$ to $0.50\,B_0$, the current center moves monotonically from $+19$ to $-18\,d_e$, whereas $\delta_{BT}$ stays within $2.0$--$2.5\,d_e$, $\delta_c$ within $2.0$--$2.9\,d_e$, and $L_{\rm EDR}$ within $7.7$--$9.3\,d_e$. Across these controls the coordinate change and the imposed offsets mainly reposition the sheet, while the local width physics analyzed in the remainder of the paper is mostly preserved.

\section{Reproduction of the MRX Electron Layer}
\label{sec:mrx_benchmark}

This section presents the direct comparison between the primary realistic-mass run and the MRX measurements. Section~\ref{sec:local_width} compares the local layer width in centimeters and in $d_e$ units. Section~\ref{sec:momentum} presents the electron force balance that supports the reconnection electric field. Section~\ref{sec:grl_accounting} applies the force-balance accounting used in the experiment to the simulation.

\subsection{Local width in centimeters and electron-inertial units}
\label{sec:local_width}

Prior MRX-type kinetic simulations at reduced mass ratio reported electron-layer half-widths of $1.5$--$3\,d_e$  \cite{Dorfman2008PoP,Roytershteyn2010PoP}, smaller than the probe-corrected experimental scale. The realistic-mass-ratio collisional run here provides a more direct local comparison against the MRX electron-layer width. For the MRX reference range we use the probe-blockage correction reported by Ji et al.~\cite{Ji2008GRL}. Probe blockage inflates the measured electron-layer half-width by a multiplicative factor of 1.06--1.44 (a 6--44\% increase), depending on the ratio of the layer thickness to the glass-tube radius, so the reported $\sim8\,d_e$ scale corresponds to a corrected range of $5.5$--$7.5\,d_e$.

Using the 40\% downstream electron-outflow definition \cite{Dorfman2008PoP,RenThesis}, the VPIC simulated half-width is $\delta_{BT}=0.744\pm0.054$ cm, or $6.26\pm0.45\,d_e$ in this realistic scenario. The centimeter value lies within the $0.4$--$1.1$ cm range implied by the MRX measurements, and the $d_e$ value lies within the probe-corrected $5.5$--$7.5\,d_e$ range, as summarized in Table~\ref{tab:metric_comparison}. Both MRX entries in Table~\ref{tab:metric_comparison} derive from Ji et al.~\cite{Ji2008GRL}: the $\delta_{BT}$ row in centimeters is the range implied by the probe-corrected measurement and the experimental electron skin depth ($5.5$--$7.5\,c/\omega_{pe}$ at $c/\omega_{pe}=0.7$--1.5 mm), and the $\delta_{BT}/d_e$ row is the probe-corrected range. The agreement in centimeters and in $d_e$ units provides the local-width comparison.

\begin{table}[!htbp]
\centering
\caption{Local MRX comparison for the realistic-mass collisional run. VPIC uncertainties are standard deviations over $t\Omega_{ci}=80$--110, and MRX entries are probe-blockage corrected ranges.}
\label{tab:metric_comparison}
\tablecaptiongap
\small
\setlength{\tabcolsep}{2.4pt}
\begin{tabular}{L{0.30\linewidth}L{0.28\linewidth}L{0.28\linewidth}}
\hline
Metric & MRX & VPIC \\
\hline
$\delta_{BT}$ [cm] & $0.4$--$1.1$ & $0.744\pm0.054$ \\
$\delta_{BT}/d_e$ & $5.5$--$7.5$ & $6.26\pm0.45$ \\
\hline
\end{tabular}
\end{table}

\subsection{Electron force balance at the X line}
\label{sec:momentum}

\begin{figure}[!t]
\centering
\includegraphics[width=\singlefigwidth]{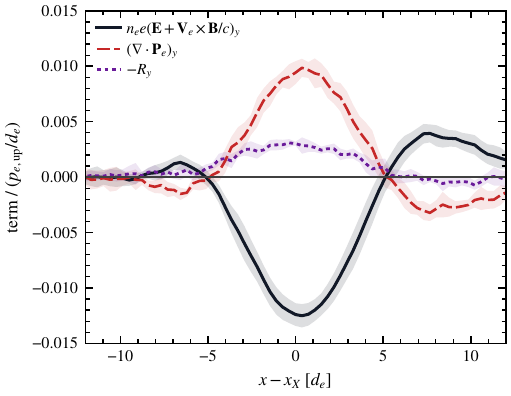}
\caption{Local out-of-plane force balance in the realistic-mass collisional run. Terms of Eq.~\eqref{eq:y_force_balance} averaged over $t\Omega_{ci}=80$--110 are plotted across the X line, and shaded bands show the standard deviation over the same interval. Terms are normalized by $p_{e,\rm up}/d_e$, where $p_{e,\rm up}$ is the scalar electron pressure at the same $3\,d_i$ upstream reference point that defines $B_0$ and $\beta_{e,0}$ in Sec.~\ref{sec:setup}, averaged over the two inflow sides of the X line. The black curve shows the nonideal term $n_ee({\bf E}+{\bf V}_e\times{\bf B}/c)_y$, and the red and purple curves show the pressure-divergence term $(\nabla\cdot{\bf P}_e)_y$ and the residual collisional channel $-R_y$. The electron time-derivative and flow-inertial terms provide at most a few percent of the support and are therefore not plotted.}
\label{fig:force}
\end{figure}

The second observable is the balance that supports the reconnection electric field. We take the out-of-plane component of the electron momentum equation and write it in force-density form, the convention plotted in Fig.~\ref{fig:force}, with ${\bf V}_e$ the electron bulk velocity,
\begin{equation}
\begin{aligned}
  &n_e\,e\left({\bf E}+\frac{{\bf V}_e\times{\bf B}}{c}\right)_y
  =
  -(\nabla\cdot{\bf P}_e)_y \\
  &\quad
  -m_e n_e
  \left(
  \partial_t V_{ey}+{\bf V}_e\cdot\nabla V_{ey}
  \right)
  +R_y.
\end{aligned}
\label{eq:y_force_balance}
\end{equation}
We refer to Eq.~\eqref{eq:y_force_balance} throughout as the electron force balance. The residual force density $R_y$ is the remainder after the electric, magnetic, pressure-divergence, time-derivative, and flow-inertial terms are evaluated. In the present reduction it serves as a collisional channel and bounds the collisional drag from above, because it absorbs all unresolved contributions. In the collisional limit, $R_y$ contains both the friction force (resistivity) and the thermal force \cite{Daughton2009PRL}. While closure relations of the Braginskii type agree well with this term in the collisional regime (e.g., Ref.~\cite{Daughton2009PRL}), such closures can break down when collisions become infrequent, or when $|E_y| > E_D$.

We summarize the division of nonideal support by the pressure-support fraction, evaluated in the same box of one-$d_e$ half-width centered on the X-line current center,
\begin{equation}
  F_P
  =
  \frac{|(\nabla\cdot{\bf P}_e)_y|}
  {|(\nabla\cdot{\bf P}_e)_y|+|R_y|}.
  \label{eq:pressure_fraction}
\end{equation}
In the collisionless limit the pressure tensor carries the entire nonideal support and $F_P\rightarrow1$. Increasing collisionality transfers support to the collisional channel.

Figure~\ref{fig:force} shows the balance for the primary realistic-mass run. The nonideal force density on the left-hand side of Eq.~\eqref{eq:y_force_balance} is supported by the pressure-divergence and residual collisional channels with $F_P\simeq0.76$ and the collisional fraction $1-F_P\simeq0.24$. In this two-dimensional picture, the force balance of Eq.~\eqref{eq:y_force_balance} closes with these two terms alone.

\subsection{The experimental force-balance accounting revisited}
\label{sec:grl_accounting}

In 2008, Ji et al.~\cite{Ji2008GRL} tested whether the classical electron-layer channels could account for the measured reconnection electric field in MRX. Near the X line the nongyrotropic pressure contribution is well approximated by the Hesse estimate \cite{Hesse1999PoP}
\begin{equation}
  E_{\rm NG}
  \equiv
  -\left(\frac{\nabla\cdot{\bf P}_e}{e n_e}\right)_y
  \approx
  \frac{1}{e}\,\frac{\partial V_{ez}}{\partial z}\,\sqrt{2 m_e T_e},
  \label{eq:hesse_estimate}
\end{equation}
and the collisional contribution by the Spitzer field $E_\eta=\eta_{\rm Spitzer} j_y$, with $\eta_{\rm Spitzer}$ the Spitzer resistivity. Evaluated with the measured MRX profiles, the two terms together accounted for only 30--45\% of the measured $E_\theta$ (the experimental toroidal field, our $E_y$), and the deficit motivated dissipation mechanisms beyond two-dimensional Coulomb-collisional dynamics.\cite{Ji2008GRL}

Applying the identical accounting inside the primary realistic-mass run gives a different outcome. Evaluating Eq.~\eqref{eq:hesse_estimate} with the X-line shear $\partial V_{ez}/\partial z$ gives $E_{\rm NG}=0.78\,|E_y|$, in agreement with the measured pressure-divergence support of $0.72$--$0.76\,|E_y|$ from Fig.~\ref{fig:force}. The Spitzer friction evaluated with the local collision rate and the unmagnetized (parallel) Spitzer coefficient, the appropriate limit at the field null, gives $0.20\,|E_y|$, in agreement with the measured residual share of $0.22\,|E_y|$ (the collisional fraction $1-F_P\simeq0.24$ expressed in units of $|E_y|$). The two classical channels close the balance.

\begin{figure*}[!t]
\centering
\includegraphics[width=\widefigwidth]{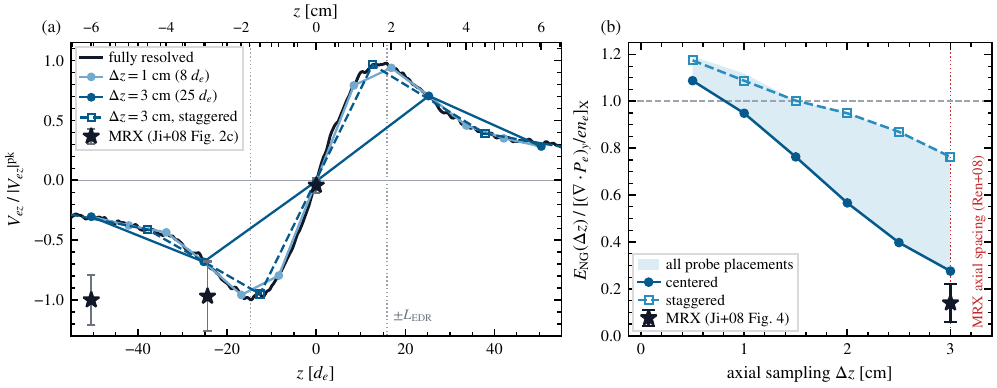}
\caption{The primary run read through the experiment's axial sampling.
(a) Fully resolved electron outflow $V_{ez}(z)$ through the X line of
the primary run (black), and the same profile as recorded by probe
arrays with axial point spacing $\Delta z=1$ and 3~cm (piecewise-linear
through the sampled points; filled circles, centered placement with
a probe on the X line; open squares, the 3~cm staggered placement
with the X line midway between probes), each curve normalized to its
peak. Dotted vertical lines mark $\pm L_{\rm EDR}$. Stars, with the read-off uncertainty as error bars, show the MRX points read from Fig.~2(c) of Ref.~\cite{Ji2008GRL} at their 3~cm stations on the $z\le0$ side, normalized to their own peak.
(b) The Hesse estimate of Eq.~\eqref{eq:hesse_estimate}, evaluated with
the finite-difference shear at spacing $\Delta z$, as a fraction of the
measured pressure-gradient support $[(\nabla\cdot{\bf P}_e)_y/en_e]_{\rm X}$.
Shading spans probe-grid placements between the centered and staggered cases, and the dashed horizontal line marks full recovery of the measured support. The black star and error bar at the 3~cm MRX spacing \cite{Ren2008PRL} show $E_{\rm NG}/(E_\theta-E_\eta)=0.06$--$0.22$
read from Fig.~4 of Ref.~\cite{Ji2008GRL}, plotted under the accounting
assumption that the pressure channel supports $E_\theta-E_\eta$.}
\label{fig:sampling_accounting}
\end{figure*}

The deficit reported from experiments can be traced quantitatively by processing the simulation data through the experiment's operational chain. The probe arrays that yield the electron-flow profiles resolve 2.5~mm in the cross-sheet direction but are spaced 3~cm along the outflow \cite{Ren2008PRL}, meaning the accessible axial derivative at the X line is a finite difference over $25$--$50\,d_e$. The electron outflow ramp rises over a few $d_e$ and peaks near $z\simeq L_{\rm EDR}$, so any measurement chord spanning a distance comparable to or larger than $L_{\rm EDR}$ inevitably underestimates the central slope. When applied to the fully resolved simulation profile, a 3~cm finite difference captures only $20\%$ to $70\%$ of the true X-line shear, depending on where the arrays sit relative to the X line [Fig.~\ref{fig:sampling_accounting}(a)]. The same down-sampling method leaves the remaining inputs essentially untouched, changing $T_e$, $n_e$, and $\delta_{BT}$ by at most a few percent. Evaluated through this observational filter, the force-balance accounting of the simulated layer yields $E_{\rm NG}+E_\eta\simeq0.35$--$0.75\,|E_y|$, overlapping the 30--45\% support reported for MRX. Figure~\ref{fig:sampling_accounting}(b) shows this accounting directly: the sampled Hesse estimate falls from the full pressure-gradient support at fine spacing to $0.28$--$0.76$ of it at the 3~cm spacing, and the values evaluated from the measured profiles \cite{Ji2008GRL} lie at the lower edge of this range.

Taken together, the fully resolved force balance and the observational sampling-chain reproduction address the force-balance side of the historical comparison: downsampling the simulation data to the resolution of the MRX probe array reduces the measured nongyrotropic contribution by an amount similar to what is measured in MRX, which suggests that the lack of force balance measured in MRX may be due to probe resolution rather than anomalous terms. Within the two-dimensional description, the balance closes without any additional term that three-dimensional physics might introduce.

\section{Orbit-Level Origin of the Semi-Collisional Layer Width}
\label{sec:model}

Section~\ref{sec:mrx_benchmark} established the reproduction at the level of profiles and moments: a layer of the measured width whose force balance closes through the classical channels. This section explains that result at the level of electron orbits. We first measure which orbit classes exist in the reversing field and which of them carries the current (Sec.~\ref{sec:vspace_complete}); the meandering class emerges as the width-setting population at every scanned collisionality. We then construct an ordering for how this population responds when Coulomb collisions interrupt orbit coherence (Sec.~\ref{sec:ordering}), and develop an analytic model predicting the layer width. The model is tested against the collisionality scan (Sec.~\ref{sec:scan_width}).

\subsection{Orbit classes and the current-carrying population}
\label{sec:vspace_complete}

\begin{figure*}[!t]
\centering
\includegraphics[width=\widefigwidth]{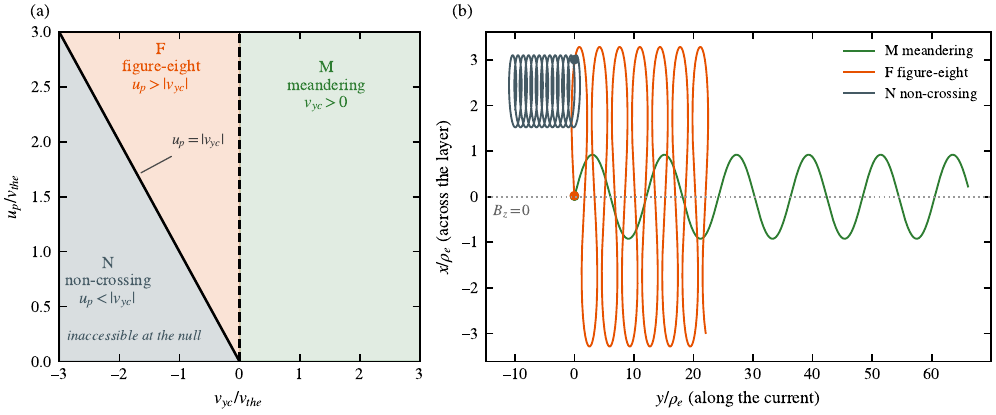}
\caption{Orbit classes in the reversing field. (a) The invariant plane spanned by the drift-frame invariant $v_{yc}$ and the speed $u_p$ perpendicular to the reversing field: every orbit is meandering (M, $v_{yc}>0$), figure-eight (F, $v_{yc}<0$, $u_p>|v_{yc}|$), or non-crossing (N, $u_p<|v_{yc}|$), the last of which closes on one side of the reversal and is therefore inaccessible at the null; the solid line is the class boundary $u_p=|v_{yc}|$. (b) One exact orbit per class in the linear reversed field, colored by class and separated along the panel, with the dotted horizontal line marking the field reversal $B_z=0$; the meandering orbit carries net $+y$ motion, the figure-eight orbit traverses with weaker $y$ coherence, and the non-crossing orbit closes on one side of the reversal.}
\label{fig:orbit_classes}
\end{figure*}

Near the X line, where the magnetic field reverses across the layer, two constants of motion organize the electron orbits in the reversing field. Conservation of the out-of-plane canonical momentum gives each electron the drift-frame invariant $v_{yc}=v_y-a(x)$, where $a(x)=(e/m_ec)\,A_y(x)$ is the canonical shift in velocity units and $A_y(x)=\int_{x_X}^{x}B_z\,dx'$ is the flux function referenced to the reversal, with the coordinate orientation chosen so that current-carrying motion has $v_{yc}>0$; for a crossing particle, $v_{yc}$ is equivalently the out-of-plane velocity when it crosses the magnetic null. In the absence of the reconnection field and collisions, the speed $u_p=(v_x^2+v_y^2)^{1/2}$ in the plane perpendicular to the reversing field is also conserved. These two constants classify every orbit, following the invariant-based orbit analysis of Refs.~\cite{Buchner1989JGR,Daughton1999PoP}, into one of three classes (Fig.~\ref{fig:orbit_classes}): meandering orbits (M, $v_{yc}>0$) that cross the reversal with net $+y$ motion, figure-eight orbits (F, $v_{yc}<0$ with $u_p>|v_{yc}|$) that traverse the reversal with weaker $y$ coherence, and non-crossing orbits (N, $u_p<|v_{yc}|$) whose excursions close on one side of the reversal and which populate the $v_y\le0$ side of the distribution. The layer contains all three, so which class controls the width is a measurable question.

\begin{figure}[!tb]
\centering
\includegraphics[width=\singlefigwidth]{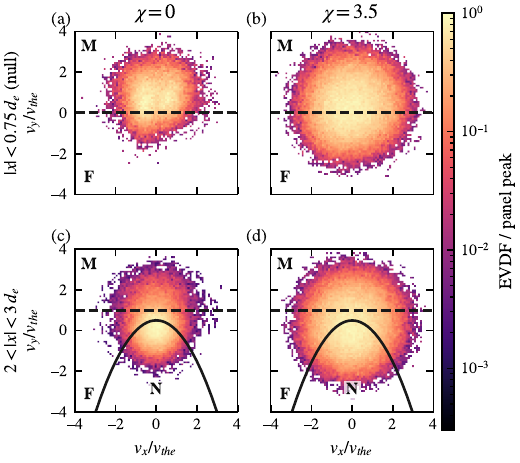}
\caption{Measured electron velocity distributions $f(v_x,v_y)$ for the $m_i/m_e=100$ particle-scan cases: at the field null ($|x|<0.75\,d_e$) in (a,b) and in the adjacent slabs $2<|x|<3\,d_e$ in (c,d), at $\chi=0$ in (a,c) and $\chi=3.5$ in (b,d), each normalized to its panel peak. The color scale is logarithmic. The black dashed line marks the meandering/figure-eight boundary at the box-averaged canonical shift $\langle a\rangle$. The black parabola in (c,d) bounds the non-crossing population N; it is absent from (a,b) because $\langle a\rangle=0$ at the null, where the non-crossing class is inaccessible.}
\label{fig:evdf_classes}
\end{figure}

To determine the role of each electron class, electron velocity distributions are recorded in boxes around the X line for eight $m_i/m_e=100$ cases spanning the Dreicer ratio $\chi=E_D/|E_y|=0$--3.5, seven members of the collisional scan together with a collisionless companion run, and each particle is classified by its measured invariants: $v_{yc}$ from the measured flux function and $u_p$ from its velocity. Figure~\ref{fig:evdf_classes} shows the measured $f(v_x,v_y)$ at the magnetic null and in the adjacent slabs at the collisionless and $\chi=3.5$ endpoints, with the class boundaries overlaid.

\begin{figure}[!tb]
\centering
\includegraphics[width=\singlefigwidth]{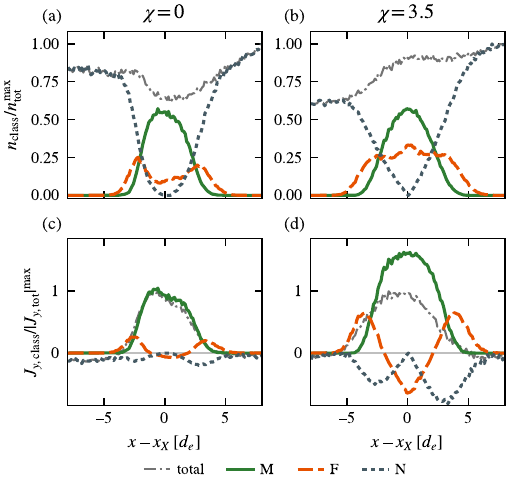}
\caption{Measured class-resolved profiles across the layer for the $m_i/m_e=100$ endpoints, at $\chi=0$ in (a,c) and $\chi=3.5$ in (b,d). In every panel the gray dash-dotted curve is the total over the three classes. (a,b) Class densities normalized to the peak total density: the meandering class peaks at the null, the non-crossing class vanishes there and dominates the wings, and the figure-eight class fills the shoulders. (c,d) Class currents normalized to the peak total current: the meandering class carries the core current, exceeding the net current at the null, with the non-crossing counter-current in the wings and, at $\chi=3.5$, a figure-eight-class return current around the null.}
\label{fig:class_profiles}
\end{figure}

The spatial distribution of each orbit class across the layer is shown in Fig.~\ref{fig:class_profiles}. The non-crossing population vanishes at the X line but dominates the outer wings: at both collisionality endpoints, the non-crossing fraction is $n_N/n_{\rm tot}<2\%$ at the reversal ($x=x_X$), while it rises past $90\%$ by $|x|\simeq5\,d_e$, where the electrons become fully magnetized. This exclusion at the null is a direct consequence of orbit geometry. Because a non-crossing orbit closes entirely on one side of the magnetic reversal, it never reaches the central null. Therefore, any electron located at the sheet center must satisfy $u_p\ge|v_{yc}|$.

The corresponding current profiles [Fig.~\ref{fig:class_profiles}(c,d)] demonstrate how these populations shape the overall current sheet. At both collisionality endpoints, the meandering class carries the primary core current. In fact, at strong collisionality ($\chi=3.5$), the meandering current exceeds the net current by a factor of 1.7. This excess is balanced by two distinct counter-currents. In the outer wings, the balance is maintained by the non-crossing population's counter-current, which corresponds to the diamagnetic response ($-\partial_x P_N/B_z$) of the magnetized electrons, with $P_N$ being the non-crossing-class pressure. Near the X-line, an additional return current emerges at higher collisionality, driven by the figure-eight class. Because figure-eight orbits ($v_{yc}<0$) traverse the null with a net $-y$ motion, their counter-current grows as collisions transfer more electrons into this class.

\begin{figure}[!tb]
\centering
\includegraphics[width=\singlefigwidth]{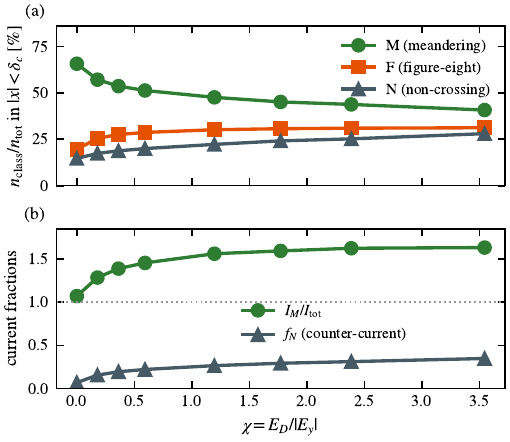}
\caption{Class-resolved population shares and currents across the $m_i/m_e=100$ particle scan, evaluated inside $|x|<\delta_c$. (a) Layer-integrated density fractions $n_{\rm class}/n_{\rm tot}$ of the meandering (M), figure-eight (F), and non-crossing (N) classes. (b) Meandering share of the net layer current $I_M/I_{\rm tot}$ and non-crossing counter-current fraction $f_N=|I_N|/I_M$; $I_M/I_{\rm tot}>1$ because the non-crossing contribution is negative; the dotted line marks $I_M/I_{\rm tot}=1$.}
\label{fig:class_census}
\end{figure}

Integrated within the layer, $|x|<\delta_c$, and across the full particle scan, the class populations shift smoothly with collisionality [Fig.~\ref{fig:class_census}(a)]: the meandering share falls from 66\% to 41\%, the figure-eight share rises from 20\% to 31\%, and the non-crossing share rises from 15\% to 28\% over $\chi=0$--3.5. The current is more selective. Writing $I_K=\int_{|x|<\delta_c}J_{y,K}\,dx$ for the current carried by class $K$, the non-crossing contribution is negative, so the meandering class carries more than the net layer current at every collisionality, $I_M/I_{\rm tot}=1.07$--1.63, with the non-crossing counter-current fraction $f_N=|I_N|/I_M$ growing from 0.07 to 0.35 across the same range [Fig.~\ref{fig:class_census}(b)]. The width-setting current is therefore carried by the meandering class throughout the scan, while the $v_{yc}\le0$ electrons enter the density and pressure balance of the layer rather than its current.

Across the scan, the meandering class remains the main current carrier: the core of the layer is dominated by its current, while the figure-eight class, whose share grows with collisionality, shapes the shoulders, where its contribution is largely offset by the non-crossing counter-current [Fig.~\ref{fig:class_profiles}(c,d)]. We note that partially cancelling class contributions of this kind could also arise if the decomposition itself becomes less physically meaningful at the highest collisionality, as the required invariance condition may break due to collisions. So we read the class-resolved attribution at $\chi=3.5$ as indicative rather than definitive. Nevertheless, the measurements narrow the target for a quantitative width model to the meandering population. Identifying the electrons whose current-carrying meandering motion survives collisions, is the subject of the next subsection.

\subsection{The coherence ordering}
\label{sec:ordering}

Section~\ref{sec:vspace_complete} identifies the meandering population as the width-setting current carrier, but the role of these orbits when Coulomb collisions interrupt their accumulation has not been characterized. We focus on this class because it carries the core current; the figure-eight and non-crossing classes enter mainly through the density and pressure balance. We therefore combine a local meandering-orbit estimate \cite{Speiser1965JGR,Sonnerup1971JGR,Buchner1989JGR} with a coherence condition applied directly in velocity space. Near the X-line current center the fields are approximated by ${\bf B}=B_H^*(x/\Delta)\hat{\bf z}$ and ${\bf E}=E_y\hat{\bf y}$, with $\Delta$ the layer half-width and $B_H^*$ the Harris-shoulder field of Eq.~\eqref{eq:harris_fit}. Here $v_x$ and $v_y$ denote the velocity components at the field reversal $x=0$, where $a(x)=0$, so the condition $v_y>0$ coincides with the meandering-class criterion $v_{yc}>0$. Electrons with $v_y>0$ execute Speiser-like bounces at the frequency $\omega_b=(\Omega_{ce}v_y/\Delta)^{1/2}$ and cross the sheet with meandering half-width $\delta_m=|v_x|/\omega_b$. Writing ${\bf u}={\bf v}/v_{the}$ and $\rho_e=v_{the}/\Omega_{ce}$, the local orbit width is
\begin{equation}
  \frac{\delta_m}{\rho_e}
  =
  \left(\frac{\Delta}{\rho_e}\right)^{1/2}\frac{|u_x|}{\sqrt{u_y}},
  \qquad u_y>0.
  \label{eq:local_meander_width}
\end{equation}

\begin{figure*}[!t]
\centering
\includegraphics[width=\widefigwidth]{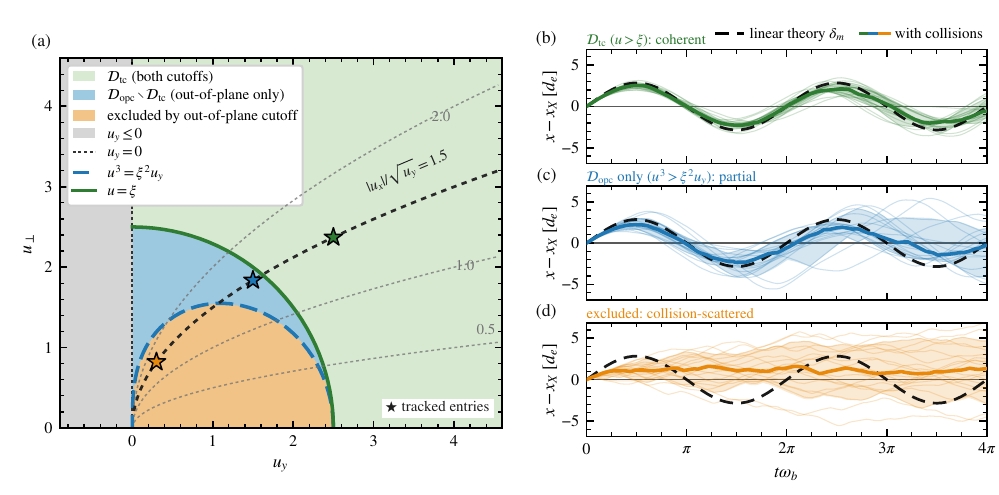}
\caption{Coherence-domain construction and tracked orbits. (a) Semi-collisional velocity-space cutoff domains in the $(u_y,u_\perp)$ half-plane, with $u_\perp=(u_x^2+u_z^2)^{1/2}$, drawn for the illustrative value $\xi=2.5$, that is $\chi=6.25$: the total-collision condition $u>\xi$ (green, $\mathcal{D}_{\rm tc}$) is nested within the out-of-plane condition $u^3>\xi^2u_y$ (blue annulus, $\mathcal{D}_{\rm opc}\setminus\mathcal{D}_{\rm tc}$); orange marks particles excluded by the out-of-plane cutoff. Dotted curves are constant meandering-amplitude loci $|u_x|/\sqrt{u_y}$, the emphasized one being $|u_x|/\sqrt{u_y}=1.5$; the stars ($u_z=0$) are one tracked entry per domain on that locus. (b)--(d) Test-particle meandering orbits for those entries, tracked with a Boris pusher and Takizuka--Abe collisions in the frozen local model field: the dashed line is the common linear-theory amplitude $\delta_m$, the heavy solid curve the collisional ensemble median, the faint curves the individual collision-seed realizations, and the shaded band their 10th--90th percentile spread. The coherent $\mathcal{D}_{\rm tc}$ entry (b) tracks the theory baseline; the excluded entry (d) is collision-scattered. Orbits use the local model field with the collision strength scaled so the coherence boundary falls near $\xi$ within the displayed two-bounce window (illustrative).}
\label{fig:cutoff_domains}
\end{figure*}

Coulomb scattering removes coherence from velocity space selectively, through the velocity-dependent rate $\nu_{ei}(v)=\nu_{ei}^{\rm th}(v_{the}/v)^3$. Comparing the electric acceleration with the collisional relaxation rate introduces the normalized speed threshold
\begin{equation}
  \xi=\left(\frac{E_D}{|E_y|}\right)^{1/2}=\chi^{1/2},
  \label{eq:dreicer_xi}
\end{equation}
with $E_D$ the Dreicer field of Eq.~\eqref{eq:ed_operational}. Two coherence criteria follow. The {\em total-collision} criterion balances the full electric force against relaxation of the full momentum, $e|{\bf E}|>\nu_{ei}(v)\,m_ev$, which reduces to $u>\xi$ near the current center where $|{\bf E}|\simeq|E_y|$. The {\em out-of-plane-collision} criterion balances $eE_y$ against relaxation of only the current-carrying component, $eE_y>\nu_{ei}(v)\,m_ev_y$, which reduces to $u^3>\xi^2u_y$. They select the nested domains
\begin{align}
  \mathcal{D}_{\rm tc}
  &=
  \left\{{\bf u}:u_y>0,\;u>\xi\right\},
  \label{eq:total_collision_cutoff}\\
  \mathcal{D}_{\rm opc}
  &=
  \left\{{\bf u}:u_y>0,\;u^3>\xi^2u_y\right\},
  \label{eq:out_of_plane_cutoff}
\end{align}
with $\mathcal{D}_{\rm tc}\subseteq\mathcal{D}_{\rm opc}$ because $0<u_y\le u$. Figure~\ref{fig:cutoff_domains} shows the domains. The green region contains particles with speeds above both cutoffs, which execute standard meandering orbits, the out-of-plane-only population (blue) retains coherent $y$-momentum even though its total speed is below the stricter threshold, and panels (b)--(d) illustrate the criterion with test-particle orbits tracked in the frozen model field, one entry per domain. In the primary run the force balance closes through the classical channels with the pressure tensor dominant, and that tensor originates at the orbit level: the off-diagonal $P_{xy}$ and $P_{yz}$ components are expected to accumulate preferentially from electrons whose $y$-momentum remains coherent through the field-reversal traversal. The Dreicer ratio is therefore the natural measure of how far this shear-carrying motion develops before collisions interrupt it, and the total-collision criterion adds the stricter requirement that the full orbit stay coherent.

A self-consistent width follows by averaging the local meandering width over a Maxwellian distribution, the turning-point averaging construction of Matsui and Daughton \cite{Matsui2008PoP,Daughton1999PoP}, restricted here to the coherent domains. The measured distributions carry a bulk drift, but in the semi-collisional regime of interest the drift is small compared with the thermal spread and is omitted for simplicity. For the isotropic normalized Maxwellian $f_M({\bf u})=(2\pi)^{-3/2}\exp(-u^2/2)$, averaging Eq.~\eqref{eq:local_meander_width} over a domain $\mathcal{D}$ gives $\langle\delta_m/\rho_e\rangle_{\mathcal{D}}=(\Delta/\rho_e)^{1/2}W_{\mathcal{D}}$, and setting the half-width equal to the selected-population orbit width gives
\begin{equation}
  \frac{\Delta_{\rm pred}}{\rho_e}
  =
  W_{\mathcal{D}}^2(\xi),
  \qquad
  W_{\mathcal{D}}(\xi)
  =
  \frac{\int_{\mathcal{D}} d^3u\,f_M({\bf u})\,|u_x|/\sqrt{u_y}}
       {\int_{\mathcal{D}} d^3u\,f_M({\bf u})}.
  \label{eq:self_consistent_width}
\end{equation}
For $\xi\rightarrow0$ both domains reduce to the collisionless meandering floor, and for $\xi\gg1$ they separate:
\begin{align}
  \frac{\Delta_{\rm tc}}{\rho_e},
  \frac{\Delta_{\rm opc}}{\rho_e}
  &\rightarrow
  1.88
  \qquad(\xi\rightarrow0),
  \label{eq:collisionless_width_floor}\\
  \frac{\Delta_{\rm tc}}{\rho_e}
  &\rightarrow
  1.24\,\xi,
  \label{eq:tc_width_asymptote}\\
  \frac{\Delta_{\rm opc}}{\rho_e}
  &\rightarrow
  0.833\,\xi^2,
  \label{eq:opc_width_asymptote}
\end{align}
with the closed-form floor and the $\xi\gg1$ coefficients derived in Appendix~\ref{app:maxwellian_derivation}. The out-of-plane asymptote, $\Delta_{\rm opc}\propto\xi^2\rho_e=\chi\,\rho_e$, recovers the linear-in-collisionality resistive-fluid current layer scaling of Roytershteyn et al.~\cite{Roytershteyn2010PoP}, which retains only the out-of-plane momentum equation. The total-collision criterion grows more slowly with collisionality because it additionally requires the cross-sheet $x$ and outflow $z$ components of the orbit to remain coherent.

\subsection{Width and force-balance partition across the collisionality scan}
\label{sec:scan_width}
\label{sec:fp_scan}

\begin{figure}[!t]
\centering
\includegraphics[width=\singlefigwidth]{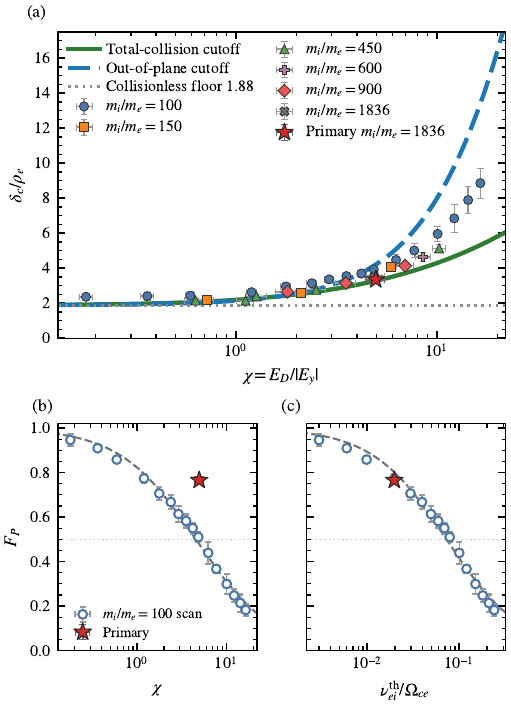}
\caption{Coherence ordering and force-balance partition across the collisional scan. (a) Measured $\delta_c/\rho_e$ as a function of $\chi=E_D/|E_y|$; the green and blue curves evaluate Eq.~\eqref{eq:self_consistent_width} over $\mathcal{D}_{\rm tc}$ and $\mathcal{D}_{\rm opc}$, and the gray dotted line marks the collisionless floor. Error bars denote variation over $t\Omega_{ci}=80$--110, points are colored and shaped by mass ratio, and the star marks the realistic-mass-ratio case. (b),(c) Measured pressure fraction $F_P$ of Eq.~\eqref{eq:pressure_fraction} for the $m_i/m_e=100$ scan members (open circles), plotted against $\chi$ in (b) and against the imposed $\nu_{ei}^{\rm th}/\Omega_{ce}$ in (c), with the primary run (star) overlaid on both. The dashed curve is a smooth monotone guide through the $m_i/m_e=100$ points, and the dotted line marks the equipartition value $F_P=0.5$.}
\label{fig:kinematic_scan}
\end{figure}

Figure~\ref{fig:kinematic_scan}(a) compares the measured $\delta_c/\rho_e$ across the collisional scan with Eq.~\eqref{eq:self_consistent_width} evaluated over $\mathcal{D}_{\rm tc}$ and $\mathcal{D}_{\rm opc}$. 

The measured widths populate the band between the two coherence curves rather than tracking either endpoint, the geometrical signature of a layer to which both populations contribute jointly: the two cutoffs bracket the measured thickness, with the stricter total-collision population setting the lower bound and the broader out-of-plane-coherent population the upper bound. The bracketing holds at moderate and strong collisionality, while the lowest-$\chi$ points ($\chi\lesssim0.6$) sit slightly above both curves, near the collisionless floor. An offset of this size is expected from the simplifications of the model, which neglects the bulk drift of the measured distributions and their departure from the isotropic Maxwellian, effects that matter most where the cutoffs are weak. The primary realistic-mass point, at $\chi=4.96\pm0.52$, follows the same pattern: propagating the local inputs through Eq.~\eqref{eq:self_consistent_width}, the total-collision and out-of-plane estimates of $4.96\pm0.19\,d_e$ and $6.36\pm0.25\,d_e$ bracket the measured $\delta_c=5.07\pm0.24\,d_e$. 

The force-balance partition [Fig.~\ref{fig:kinematic_scan}(b)] links the orbit-level picture to the moment-level one. As collisions remove coherent orbits from the domains of Eqs.~\eqref{eq:total_collision_cutoff}--\eqref{eq:out_of_plane_cutoff}, the support of $E_y$ shifts from the nongyrotropic pressure of those orbits to the collisional channel, and the measured pressure-tensor fraction $F_P$ falls accordingly across the scan. 

The scan yields one further insight: the width and the partition are organized by different variables. Width data from all six mass ratios follow a single trend in $\chi = E_D/|E_y|$ [Fig.~\ref{fig:kinematic_scan}(a)], as expected if the width is set by the coherence of meandering motion against the reconnection field, with the mass ratio entering only through $\chi$ itself. The partition is not organized by $\chi$. Plotted against $\chi$, the realistic-mass run lies well above the $m_i/m_e=100$ trend; plotted against the imposed collision rate $\nu_{ei}^{\rm th}/\Omega_{ce}$, it joins that trend [Figs.~\ref{fig:kinematic_scan}(b) and (c)]. The two variables are tied by the mass ratio, since the reconnection electric field scales with the ion Alfv\'en speed: $\chi \propto (\nu_{ei}^{\rm th}/\Omega_{ce})\sqrt{m_i/m_e}$, up to order-unity factors of $\beta_e$ and the normalized reconnection rate. The realistic-mass run and the $m_i/m_e=100$ case at the same $\nu_{ei}^{\rm th}/\Omega_{ce}=0.0197$ therefore give nearly the same partition, $F_P\simeq0.76$ and $0.77$, while sitting a factor of four apart in $\chi$, 4.96 versus 1.20, close to the predicted $\sqrt{1836/100}\simeq4.28$.

This separation shows the significance of simulation at realistic mass ratio. A reduced-mass simulation can be placed at the experimental $\chi$ or at the experimental $\nu_{ei}^{\rm th}/\Omega_{ce}$, but not at both. Matching $\chi$ raises $\nu_{ei}^{\rm th}/\Omega_{ce}$ by $\sqrt{1836/(m_i/m_e)}$ and drives the partition toward the collision-dominated side, reaching $F_P\simeq0.5$ at $m_i/m_e=100$; matching $\nu_{ei}^{\rm th}/\Omega_{ce}$ instead leaves $\chi$ smaller by the same factor, placing the layer too close to the collisionless floor. Only at $m_i/m_e=1836$ do the two line up with the experiment at once, and this is where the matching $d_e$-scale width could emerge that earlier reduced-mass studies did not obtain (Sec.~\ref{sec:local_width}).

\section{Discussion}
\label{sec:discussion}
Sections~\ref{sec:mrx_benchmark} and \ref{sec:model} presented the reproduction of the $d_e$-scale layer width, its orbit-level origin and a possible resolution of the reported force-balance deficit. This section turns to what does not yet match: the local thermodynamic state behind the $\rho_e$ normalization, and the directions, in simulation and in measurement, that could resolve it.

\subsection{Thermodynamic states and $\rho_e$ normalization}
\label{sec:thermo_norm}
\label{sec:cooling_control}

Expressing the local width in $\rho_e$ units introduces the local thermodynamic state, and Table~\ref{tab:thermo_comparison} collects this part of the comparison. With the conventions of Sec.~\ref{sec:diagnostics} and Eq.~\eqref{eq:rhoe_beta_relation}, the MRX profiles give $\beta_e\simeq0.34$--0.55 and $\rho_e/d_e\simeq0.42$--0.53, while the quasi-steady VPIC X line is depleted and heated to $\beta_e\simeq2.95$--3.46 and $\rho_e/d_e\simeq1.49$--1.54, so the same simulated layer maps to very different $\delta/\rho_e$ values while the centimeter and $d_e$ comparisons of Table~\ref{tab:metric_comparison} agree. The measured width exceeds the simulated one by a factor of 3.1--4.3 in $\rho_e$ units, while the MRX $\rho_e/d_e$ is smaller than the simulated value by a factor of 2.8--3.6, so the product, the width in $d_e$, agrees.

\begin{table}[!htbp]
\centering
\caption{Thermodynamic-state comparison behind the $\rho_e$-normalized width. The MRX rows derive from the quasi-steady point set of Refs.~\cite{Ji2008GRL,Roytershteyn2010PoP}: the width row is probe-blockage corrected (5--45\% envelope), and the thermodynamic rows are interquartile ranges.}
\label{tab:thermo_comparison}
\tablecaptiongap
\small
\setlength{\tabcolsep}{2.4pt}
\begin{tabular}{L{0.30\linewidth}L{0.28\linewidth}L{0.28\linewidth}}
\hline
Metric & MRX & VPIC \\
\hline
$\delta_c/\rho_e$ & $10.4$--$14.4$ & $3.35\pm0.12$ \\
$\rho_e/d_e$ & $0.42$--$0.53$ & $1.49$--$1.54$ \\
$\beta_e$ & $0.34$--$0.55$ & $2.95$--$3.46$ \\
\hline
\end{tabular}
\end{table}
The MRX discharges of Ref.~\cite{Roytershteyn2010PoP} lie in the same collisionality range as the present simulations: their Dreicer ratios are $\chi\simeq1$--10 ($E_y/E_{\rm crit}=0.1$--1 in their notation), so the collisionality imposed here is representative of the experiment. At these $\chi$, the ordering discussed above predicts widths of a few $\rho_e$ and every simulation in the scan measures $2.2$--$3.4\,\rho_e$, whereas the MRX widths are a factor of 3.9--8.4 larger at each point's own $\chi$ (2.7--5.8 after the maximal blockage correction). Ref.~\cite{Roytershteyn2010PoP} read this excess as evidence for physical mechanisms beyond two-dimensional Coulomb dynamics, with lower-hybrid fluctuations the leading candidate \cite{Ji2004PRL,Carter2001PRL}, although the three-dimensional simulations of Ref.~\cite{Roytershteyn2013PoP} found the resulting broadening insufficient at the parameters tested. Here we examine a second possibility. The normalizing scale $\rho_e$ depends on the local thermodynamic state of the layer, which the simulations do not reproduce (Table~\ref{tab:thermo_comparison}); a mismatch in that state would enter the $\rho_e$-normalized width directly, independently of additional scattering. Since the $d_e$-scale width already matches the experiment and the force balance of Sec.~\ref{sec:mrx_benchmark} closes within two-dimensional Coulomb dynamics, this possibility merits examination.

\begin{figure}[!t]
\centering
\includegraphics[width=\singlefigwidth]{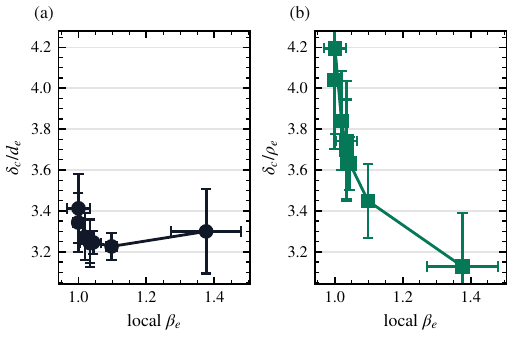}
\caption{Cooling-control response for the $m_i/m_e=100$ sequence: (a) $\delta_c/d_e$ and (b) $\delta_c/\rho_e$, plotted versus the local $\beta_e$ of Eq.~\eqref{eq:rhoe_beta_relation}, from the uncooled baseline (the largest-$\beta_e$ point) toward the strongest cooling. Error bars are standard deviations over $t\Omega_{ci}=80$--110, and the line connects the sequence in order of increasing cooling. This reduced-mass sequence sits at lower $\beta_e$ than the realistic-mass primary run of Table~\ref{tab:thermo_comparison}.}
\label{fig:beta_m100}
\end{figure}

Table~\ref{tab:thermo_comparison} shows the two mismatches to be of the same size and opposite sign, so that the $d_e$-scale comparison agrees while the $\rho_e$-scale one does not. Whether that compensation is structural or a coincidence of two unrelated errors can be tested directly, by moving the thermodynamic state and watching both normalizations respond. We therefore introduce an additional electron energy sink. This cooling-control sequence (Table~\ref{tab:run_parameters}) applies an ad hoc $\propto T_e^4$ heat sink to all electron macroparticles every five VPIC time steps, at fixed reconnection drive and fixed imposed collisionality, standing in for the radiative and transport losses present in the experiment. Between the baseline and the strongest cooling point the central temperature drops by nearly a factor of two and the local $\beta_e$ falls from 1.4 to 1.0. Figure~\ref{fig:beta_m100} shows the response: the $d_e$-normalized width holds to within $\sim3\%$, while the $\rho_e$-normalized width rises by about 34\%. 

The sequence does not, however, close the gap. First, the control run saturates near $\beta_e\simeq1.0$ and does not reach the sub-unity $\beta_e$ of the experiment, a structural constraint of the uniform-fill initial condition, for which pressure balance across the layer keeps the local $\beta_e$ near or above unity; in the experiment, elements not modeled here, such as a non-uniform background plasma profile and a weak self-generated guide field contributing to the force balance, may have reduced $\beta_e$ further. Second, the simplified cooling term changes not only $\beta_e$ and $\rho_e$ but also the local collisionality, since the reduced central temperature makes Coulomb collisions more frequent ($\propto T_e^{-3/2}$), so the sequence fixes the direction of the response rather than the path the experiment takes to the same state. Whether a purely thermodynamic response under two-dimensional Coulomb dynamics can account for the full remaining factor is therefore left open in this simple treatment, although extrapolating the trajectory suggests that reaching $\beta_e\simeq0.34$--$0.55$ would increase the normalized width further. What the sequence does establish is that the $d_e$-scale agreement survives a change in thermodynamic state that moves the $\rho_e$-normalized width, so the two normalizations are not locked together. Narrowing the remaining difference will require more realistic models of radiative losses, plasma-neutral collisions, or three-dimensional physics such as stochastic magnetic fields that conduct heat from the sheet center, and may also require measurement-side refinements in estimating the local $\rho_e$ in MRX.

\subsection{Remaining open items and outlook}
\label{sec:one_anomaly}

Of the two historical discrepancies, Sec.~\ref{sec:grl_accounting} suggests that the measured force-balance deficit may be a result of insufficient probe resolution in the axial ($Z$) direction. However, the $\rho_e$-normalized width and the exact thermodynamic state remain open, demanding further effort on both the simulation and measurement fronts.

On the simulation side, the physical mechanisms that plausibly establish the experimental thermodynamic state are not yet fully modeled. Several critical factors remain missing: (1) three-dimensional electron heat transport along non-axisymmetric, or even stochastic, magnetic fields, with the classical parallel heat diffusivity $\kappa_\parallel\simeq3.2\,v_{the}^2\tau_e$, where $\tau_e$ is the electron collision time \cite{Braginskii1965}, which could drain the layer across the $2L\simeq40$~cm flux-core separation on a timescale ($\tau_\parallel\sim L^2/\kappa_\parallel\approx1~\mu$s) far shorter than the drive; (2) electron-neutral collisions and radiation in the partially ionized discharge; (3) downstream pressure buildup in the confined outflow, for which no direct experimental measurement is currently available for comparison; and (4) the push-phase formation history, which sets the initial profiles for the pull phase, unlike the uniform fill used for the initial conditions in this study. Especially, the $\rho_e$ normalization itself combines quantities from different locations, the electron temperature at the layer center with the Harris-shoulder field $B_H^*$ upstream. So reproducing its experimental value requires modeling the cross-layer profiles rather than the X-line state alone; the electron--neutral and radiative channels of item (2) act directly on those profiles.

On the measurement side, the axial probe resolution that Sec.~\ref{sec:grl_accounting} identifies bears also on the layer length. The simulated EDR half-length, $L_{\rm EDR}=16.3\pm2.1\,d_e$, lies well below the $40$--$80\,d_e$ reported for MRX \cite{Ren2008PRL,Ren2008PoP}, and this is not independent of the accounting demonstrated there on the simulated $\simeq15\,d_e$ outflow ramp. Both the reported length and the underestimate of $E_{\rm NG}$ follow from sampling an outflow ramp at finite probe spacing in $Z$: a ramp shorter than the spacing is registered as a gentler gradient extended over more channels, which lengthens the inferred $L_{\rm EDR}$ while lowering the peak nonideal field recovered from the same profiles. A shorter experimental ramp, closer to the simulated value, would then be consistent with both observations at once, and the length mismatch would not constitute a separate discrepancy. Distinguishing this from a genuinely longer experimental layer requires axial resolution finer than the present probe spacing, as does the local thermodynamic state on which the $\rho_e$ normalization rests; electron-layer measurements with finer and less perturbative diagnostics would complete the picture from the other side.

\section{Conclusions}
\label{sec:conclusions}

Two-dimensional particle-in-cell simulations with binary Coulomb collisions, in a cylindrical flux-core geometry modeled on MRX and advanced to the realistic mass ratio $m_i/m_e=1836$ at MRX-relevant collisionality, reproduce the measured local electron current sheet width in absolute units and in electron-inertial units. The half-width $\delta_{BT}=0.744\pm0.054$ cm, or $6.26\pm0.45\,d_e$, lies within the experimental range in centimeters and within the probe-corrected range in $d_e$, where prior reduced-mass simulations obtained $1.5$--$3\,d_e$. The electron force balance closes through the classical channels, with the pressure-tensor divergence supporting 76\% of the nonideal electric field and collisional friction the remainder. Applying the force-balance accounting of Ji et al.~\cite{Ji2008GRL} inside the simulation shows that the Hesse nongyrotropic estimate reproduces the measured pressure support when evaluated with the gradients downsampled to the experimental resolution, so the classical channels are not intrinsically deficient in this regime.

To identify which electrons set the width, we formulated an analytic model in which the Dreicer ratio $E_D/|E_y|$ segments velocity space into nested domains that remain coherent over meandering orbits. The two coherence criteria yield distinct semi-collisional asymptotes and a common collisionless meandering floor, with a smooth transition between the two limits as $E_D/|E_y|$ grows. Across the VPIC scan, the measured widths are organized by the two criteria, bracketed by them at moderate and stronger collisionality.

Of the two historical discrepancies, what survives is the $\rho_e$-normalized width, a factor of three to six above the coherence curves evaluated at MRX's own Dreicer ratios and above every simulation in the scan, together with the local thermodynamic state, whose sub-unity $\beta_e$ no simulation in this study reaches. Introducing an electron-cooling term showed that the $d_e$-scale width persists while the $\rho_e$-scale width moves toward the measured value, though elements of the experiment not modeled here leave the gap open. Closing it will require simulations that add three-dimensional electron heat transport, electron-neutral collisions and radiation, downstream pressure buildup, and the push phase, in parallel with revisited electron-layer measurements. Both open directions converge on further collaborative investigation at the Facility for Laboratory Reconnection Experiments (FLARE) \cite{Ji2026FLARE}, which combines larger system size and higher Lundquist number, access to multiple collisionality regimes of the reconnection phase diagram \cite{JiDaughton2011PoP,Ji2022NRP}, and room for finer,
more advanced electron-layer diagnostics. 

\begin{acknowledgments}
This work was supported by the U.S. Department of Energy under contract \# DE-AC02-09CH11466. T.Rhee was supported by the R\&D Program of the Korea Institute of Fusion Energy (KFE), which is funded by the Ministry of Science and ICT of the Republic of Korea (KFE-EN254). Part of T.Rhee's work was also supported by the National Aeronautics and Space Administration (NASA) under \#80HQTR21T0105.
\end{acknowledgments}

\section*{Author Declarations}
\subsection*{Conflict of Interest}
The authors have no conflicts to disclose.

\section*{Data Availability}
The data that support the findings of this study are available from the corresponding author upon reasonable request.

\appendix
\setcounter{figure}{0}
\renewcommand{\thefigure}{A\arabic{figure}}
\renewcommand{\theHfigure}{appendix.\arabic{figure}}

\section{Fit and Scan Diagnostics}
\label{app:fit_width}

\begin{figure}[!htb]
\centering
\includegraphics[width=\singlefigwidth]{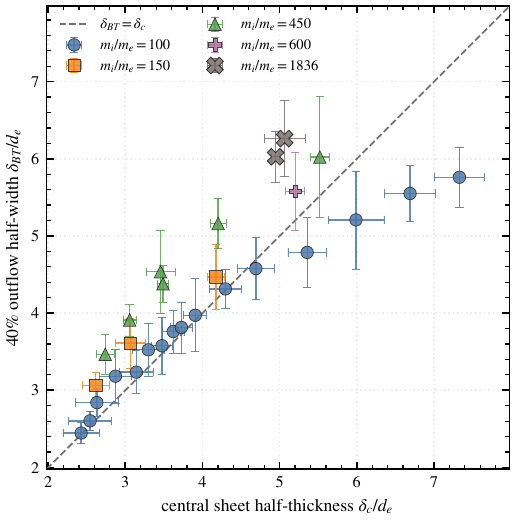}
\caption{Relation between $\delta_{BT}$ and $\delta_c$ in the collisional scan. Points show time-window means, and error bars show standard deviations. Colors and markers follow the mass-ratio key of Fig.~\ref{fig:kinematic_scan}, and the dashed line denotes $\delta_{BT}=\delta_c$.}
\label{fig:sheetwidth}
\end{figure}

Figure~\ref{fig:sheetwidth} compares $\delta_{BT}$ and $\delta_c$ in the collisional scan. The comparison includes all scan subsets used in the main theory plot, so the relation should be read as a scan-level trend rather than a filtered subset. Across the scan the two widths differ by at most 31\% (median 10\%), and within the $m_i/m_e>100$ cases alone the maximum is the same 31\% (median 22\%), so the correspondence between the two definitions holds across mass ratio rather than being set by the $m_i/m_e=100$ family.

\section{Maxwellian Integrals for the Coherence-Domain Widths}
\label{app:maxwellian_derivation}
\setcounter{figure}{0}
\renewcommand{\thefigure}{B\arabic{figure}}
\renewcommand{\theHfigure}{appendixB.\arabic{figure}}
\setcounter{equation}{0}
\renewcommand{\theequation}{B\arabic{equation}}
\renewcommand{\theHequation}{appendixB.\arabic{equation}}

This appendix records the Maxwellian moment integrals that produce the collisionless floor of Eq.~\eqref{eq:collisionless_width_floor} and the semi-collisional asymptotes of Eqs.~\eqref{eq:tc_width_asymptote}--\eqref{eq:opc_width_asymptote}. We use the dimensionless velocity ${\bf u}={\bf v}/v_{the}$ and the isotropic Maxwellian $f_M({\bf u})=(2\pi)^{-3/2}\exp(-u^2/2)$.

The cutoff domains of Eqs.~\eqref{eq:total_collision_cutoff}--\eqref{eq:out_of_plane_cutoff} are most compactly written in spherical coordinates with the polar axis aligned to $\hat{\bf y}$. Writing $\tau=u_y/u\in(0,1]$ for the upper-hemisphere cosine and $\varphi$ for the azimuth in the $(x,z)$ plane,
\begin{equation}
\begin{aligned}
  u_y &= u\tau,\qquad
  u_x = u\sqrt{1-\tau^2}\cos\varphi, \\
  d^3u &= u^2\,du\,d\tau\,d\varphi,
\end{aligned}
\end{equation}
with the Speiser half-plane $\mathcal{D}_S=\{u_y>0\}$ covered by $\tau\in(0,1]$, $\varphi\in[0,2\pi)$. The local orbit-width integrand factorizes,
\begin{equation}
  \frac{|u_x|}{\sqrt{u_y}}=\sqrt{\frac{u}{\tau}}\,\sqrt{1-\tau^2}\,|\cos\varphi|,
\end{equation}
and the cutoff conditions become $u>\xi$ (total-collision) and $u>\xi\sqrt{\tau}$ (out-of-plane).

Two ingredients enter every subsequent calculation. The radial moment, obtained by the substitution $v=u^2/2$, is
\begin{equation}
  \int_a^{\infty} u^n e^{-u^2/2}\,du
  =
  2^{(n-1)/2}\,\Gamma\!\left(\frac{n+1}{2},\frac{a^2}{2}\right),
  \label{eq:radial_moment_identity}
\end{equation}
with $\Gamma(s,x)$ the upper incomplete gamma function. The angular integrals are
\begin{equation}
\begin{aligned}
  \int_0^{2\pi}|\cos\varphi|\,d\varphi &= 4, \\
  \int_0^1\tau^{-1/2}\sqrt{1-\tau^2}\,d\tau &= \frac{\Gamma(1/4)^2}{3\sqrt{2\pi}},
\end{aligned}
\label{eq:angular_integrals}
\end{equation}
the second of which follows from a standard Beta-function integral together with the reflection $\Gamma(1/4)\Gamma(3/4)=\pi\sqrt{2}$.

\textit{Collisionless floor.} For $\xi=0$, both cutoff conditions are vacuous and the average is over $\mathcal{D}_S$, with denominator $\int_{\mathcal{D}_S}f_M\,d^3u=1/2$. Using Eqs.~\eqref{eq:radial_moment_identity} and \eqref{eq:angular_integrals},
\begin{equation}
  W_{\mathcal{D}_S}
  =
  \frac{2^{1/4}\,\Gamma(1/4)}{\pi},
\end{equation}
and the self-consistent floor of Eq.~\eqref{eq:self_consistent_width} is
\begin{equation}
  \left.\frac{\Delta}{\rho_e}\right|_{\xi=0}
  =
  W_{\mathcal{D}_S}^2
  =
  \frac{\sqrt{2}\,\Gamma(1/4)^2}{\pi^2}
  \simeq1.883,
  \label{eq:floor_closed_form}
\end{equation}
yielding the value quoted in Eq.~\eqref{eq:collisionless_width_floor}.

\textit{Total-collision asymptote ($\xi\gg1$).} The cutoff $u>\xi$ is $\tau$-independent, so the radial and angular integrals separate. The leading-order asymptotic $\Gamma(s,x)\simeq x^{s-1}e^{-x}$ for large $x$ cancels the exponential factor between numerator and denominator, leaving
\begin{equation}
  \frac{\int_{\xi}^{\infty} u^{5/2}e^{-u^2/2}du}{\int_{\xi}^{\infty} u^{2}e^{-u^2/2}du}
  \xrightarrow{\xi\to\infty}\xi^{1/2},
\end{equation}
so that, combining with the angular integrals,
\begin{equation}
  W_{\mathcal{D}_{\rm tc}}(\xi)
  \xrightarrow{\xi\to\infty}
  \frac{2\,\Gamma(1/4)^2}{3\pi\sqrt{2\pi}}\,\xi^{1/2},
\end{equation}
and
\begin{equation}
  \frac{\Delta_{\rm tc}}{\rho_e}
  \xrightarrow{\xi\to\infty}
  \frac{2\,\Gamma(1/4)^4}{9\pi^3}\,\xi
  \simeq1.238\,\xi,
\end{equation}
matching Eq.~\eqref{eq:tc_width_asymptote}.

\textit{Out-of-plane asymptote ($\xi\gg1$).} The cutoff $u>\xi\sqrt{\tau}$ couples the radial and angular integrals. The substitution $\sigma=\xi^2\tau/2$ extracts the $\xi$ scaling from the $\tau$ integral via
\begin{equation}
  \int_0^{\infty}\sigma^{a-1}\,\Gamma(s,\sigma)\,d\sigma
  =
  \frac{\Gamma(s+a)}{a}.
\end{equation}
Applied to the numerator (with $s=7/4$, $a=1/2$) and the denominator (with $s=3/2$, $a=1$), the ratio yields
\begin{equation}
  W_{\mathcal{D}_{\rm opc}}(\xi)
  \xrightarrow{\xi\to\infty}
  \frac{2^{15/4}\,\Gamma(9/4)}{3\pi^{3/2}}\,\xi,
\end{equation}
so that
\begin{equation}
  \frac{\Delta_{\rm opc}}{\rho_e}
  \xrightarrow{\xi\to\infty}
  \frac{2^{15/2}\,\Gamma(9/4)^2}{9\pi^3}\,\xi^2
  \simeq0.833\,\xi^2,
\end{equation}
matching Eq.~\eqref{eq:opc_width_asymptote}.


\section*{References}
\begin{thebibliography}{99}

\bibitem{Burch2016SSR}
J. L. Burch, T. E. Moore, R. B. Torbert, and B. L. Giles, ``Magnetospheric Multiscale overview and science objectives,'' Space Sci. Rev. \textbf{199}, 5 (2016), doi:10.1007/s11214-015-0164-9.

\bibitem{Hesse1999PoP}
M. Hesse, K. Schindler, J. Birn, and M. Kuznetsova, ``The diffusion region in collisionless magnetic reconnection,'' Phys. Plasmas \textbf{6}, 1781 (1999), doi:10.1063/1.873436.

\bibitem{Kuznetsova2001JGR}
M. M. Kuznetsova, M. Hesse, and D. Winske, ``Collisionless reconnection supported by nongyrotropic pressure effects in hybrid and particle simulations,'' J. Geophys. Res. \textbf{106}, 3799 (2001), doi:10.1029/1999JA001003.

\bibitem{Speiser1965JGR}
T. W. Speiser, ``Particle trajectories in model current sheets: 1. Analytical solutions,'' J. Geophys. Res. \textbf{70}, 4219 (1965), doi:10.1029/JZ070i017p04219.

\bibitem{Sonnerup1971JGR}
B. U. \"O. Sonnerup, ``Adiabatic particle orbits in a magnetic null sheet,'' J. Geophys. Res. \textbf{76}, 8211 (1971), doi:10.1029/JA076i034p08211.

\bibitem{Buchner1989JGR}
J. B\"uchner and L. M. Zelenyi, ``Regular and chaotic charged particle motion in magnetotaillike field reversals: 1. Basic theory of trapped motion,'' J. Geophys. Res. \textbf{94}, 11821 (1989), doi:10.1029/JA094iA09p11821.

\bibitem{Horiuchi1994PoP}
R. Horiuchi and T. Sato, ``Particle simulation study of driven magnetic reconnection in a collisionless plasma,'' Phys. Plasmas \textbf{1}, 3587 (1994), doi:10.1063/1.870894.

\bibitem{Le2013PRL}
A. Le, J. Egedal, O. Ohia, W. Daughton, H. Karimabadi, and V. S. Lukin, ``Regimes of the electron diffusion region in magnetic reconnection,'' Phys. Rev. Lett. \textbf{110}, 135004 (2013), doi:10.1103/PhysRevLett.110.135004.

\bibitem{Egedal2023PoP}
J. Egedal, H. Gurram, S. Greess, W. Daughton, A. Lê, "The force balance of electrons during kinetic anti-parallel magnetic reconnection," Phys. Plasmas 30, 062106 (2023), doi:10.1063/5.0130417.

\bibitem{Egedal2024GRL}
J. Egedal, ``The adiabatic 1D kinetic equilibrium of the electron diffusion region during anti-parallel magnetic reconnection,'' Geophys. Res. Lett. \textbf{51}, e2024GL108895 (2024), doi:10.1029/2024GL108895.

\bibitem{Yamada1997MRX}
M. Yamada, H. Ji, S. Hsu, T. Carter, R. Kulsrud, N. Bretz, F. Jobes, Y. Ono, and F. Perkins, ``Study of driven magnetic reconnection in a laboratory plasma,'' Phys. Plasmas \textbf{4}, 1936 (1997), doi:10.1063/1.872336.

\bibitem{Ren2008PRL}
Y. Ren, M. Yamada, H. Ji, S. P. Gerhardt, and R. Kulsrud, ``Identification of the electron-diffusion region during magnetic reconnection in a laboratory plasma,'' Phys. Rev. Lett. \textbf{101}, 085003 (2008), doi:10.1103/PhysRevLett.101.085003.

\bibitem{Ji2008GRL}
H. Ji, Y. Ren, M. Yamada, S. Dorfman, W. Daughton, and S. P. Gerhardt, ``New insights into dissipation in the electron layer during magnetic reconnection,'' Geophys. Res. Lett. \textbf{35}, L13106 (2008), doi:10.1029/2008GL034538.

\bibitem{Dorfman2008PoP}
S. Dorfman, W. Daughton, V. Roytershteyn, H. Ji, Y. Ren, and M. Yamada, ``Two-dimensional fully kinetic simulations of driven magnetic reconnection with boundary conditions relevant to the Magnetic Reconnection Experiment,'' Phys. Plasmas \textbf{15}, 102107 (2008), doi:10.1063/1.2991361.

\bibitem{Roytershteyn2010PoP}
V. Roytershteyn, W. Daughton, S. Dorfman, Y. Ren, H. Ji, M. Yamada, H. Karimabadi, L. Yin, B. J. Albright, and K. J. Bowers, ``Driven magnetic reconnection near the Dreicer limit,'' Phys. Plasmas \textbf{17}, 055706 (2010), doi:10.1063/1.3399787.

\bibitem{Ji2004PRL}
H. Ji, S. Terry, M. Yamada, R. Kulsrud, A. Kuritsyn, and Y. Ren, ``Electromagnetic fluctuations during fast reconnection in a laboratory plasma,'' Phys. Rev. Lett. \textbf{92}, 115001 (2004), doi:10.1103/PhysRevLett.92.115001.

\bibitem{Carter2001PRL}
T. A. Carter, H. Ji, F. Trintchouk, M. Yamada, and R. M. Kulsrud, ``Measurement of lower-hybrid drift turbulence in a reconnecting current sheet,'' Phys. Rev. Lett. \textbf{88}, 015001 (2001), doi:10.1103/PhysRevLett.88.015001.

\bibitem{Daughton2003PoP}
W. Daughton, ``Electromagnetic properties of the lower-hybrid drift instability in a thin current sheet,'' Phys. Plasmas \textbf{10}, 3103 (2003), doi:10.1063/1.1594724.

\bibitem{Roytershteyn2013PoP}
V. Roytershteyn, S. Dorfman, W. Daughton, H. Ji, M. Yamada, and H. Karimabadi, ``Electromagnetic instability of thin reconnection layers: comparison of three-dimensional simulations with MRX observations,'' Phys. Plasmas \textbf{20}, 061212 (2013), doi:10.1063/1.4811371.

\bibitem{Dreicer1959PR}
H. Dreicer, ``Electron and ion runaway in a fully ionized gas. I,'' Phys. Rev. \textbf{115}, 238 (1959), doi:10.1103/PhysRev.115.238.

\bibitem{Daughton2009PRL}
W. Daughton, V. Roytershteyn, B. J. Albright, H. Karimabadi, L. Yin, and K. J. Bowers, ``Transition from collisional to kinetic regimes in large-scale reconnection layers,'' Phys. Rev. Lett. \textbf{103}, 065004 (2009), doi:10.1103/PhysRevLett.103.065004.

\bibitem{JaraAlmonte2021PRL}
J. Jara-Almonte and H. Ji, ``Thermodynamic phase transition in magnetic reconnection,'' Phys. Rev. Lett. \textbf{127}, 055102 (2021), doi:10.1103/PhysRevLett.127.055102.

\bibitem{Zocco2011PoP}
A. Zocco and A. A. Schekochihin, ``Reduced fluid-kinetic equations for low-frequency dynamics, magnetic reconnection, and electron heating in low-beta plasmas,'' Phys. Plasmas \textbf{18}, 102309 (2011), doi:10.1063/1.3628639.

\bibitem{Loureiro2013PRL}
N. F. Loureiro, A. A. Schekochihin, and A. Zocco, ``Fast collisionless reconnection and electron heating in strongly magnetized plasmas,'' Phys. Rev. Lett. \textbf{111}, 025002 (2013), doi:10.1103/PhysRevLett.111.025002.

\bibitem{Stanier2015PoP}
A. Stanier, A. N. Simakov, L. Chac\'on, and W. Daughton, ``Fluid vs. kinetic magnetic reconnection with strong guide fields,'' Phys. Plasmas \textbf{22}, 101203 (2015), doi:10.1063/1.4932330.

\bibitem{Matsui2008PoP}
T. Matsui and W. Daughton, ``Kinetic theory and simulation of collisionless tearing in bifurcated current sheets,'' Phys. Plasmas \textbf{15}, 012901 (2008), doi:10.1063/1.2832679.

\bibitem{Bowers2008PoP}
K. J. Bowers, B. J. Albright, L. Yin, B. Bergen, and T. J. T. Kwan, ``Ultrahigh performance three-dimensional electromagnetic relativistic kinetic plasma simulation,'' Phys. Plasmas \textbf{15}, 055703 (2008), doi:10.1063/1.2840133.

\bibitem{VPIC}
K. J. Bowers, B. J. Albright, L. Yin, W. Daughton, V. Roytershteyn, B. Bergen, and T. J. T. Kwan, ``Advances in petascale kinetic plasma simulation with VPIC and Roadrunner,'' J. Phys.: Conf. Ser. \textbf{180}, 012055 (2009), doi:10.1088/1742-6596/180/1/012055.

\bibitem{Greess2021JGR}
S. Greess, J. Egedal, A. Stanier, W. Daughton, J. Olson, A. Lê, R. Myers, A. Millet-Ayala, M. Clark, J. Wallace, D. Endrizzi, and C. Forest, ``Laboratory verification of electron-scale reconnection regions modulated by a three-dimensional instability,'' J. Geophys. Res.: Space Phys. \textbf{126}, e2021JA029316 (2021), doi:10.1029/2021JA029316.

\bibitem{Greess2022PoP}
S. Greess, J. Egedal, A. Stanier, J. Olson, W. Daughton, A. Lê, A. Millet-Ayala, C. Kuchta, and C. B. Forest, ``Kinetic simulations verifying reconnection rates measured in the laboratory, spanning the ion-coupled to near electron-only regimes,'' Phys. Plasmas \textbf{29}, 102103 (2022), doi:10.1063/5.0101006.

\bibitem{Takizuka1977JCP}
T. Takizuka and H. Abe, ``A binary collision model for plasma simulation with a particle code,'' J. Comput. Phys. \textbf{25}, 205 (1977), doi:10.1016/0021-9991(77)90099-7.

\bibitem{RenThesis}
Y. Ren, Ph.D. thesis, Princeton University (2008).

\bibitem{Daughton1999PoP}
W. Daughton, ``The unstable eigenmodes of a neutral sheet,'' Phys. Plasmas \textbf{6}, 1329 (1999), doi:10.1063/1.873374.

\bibitem{Braginskii1965}
S. I. Braginskii, ``Transport processes in a plasma,'' in \textit{Reviews of Plasma Physics}, edited by M. A. Leontovich (Consultants Bureau, New York, 1965), Vol. 1, p. 205.

\bibitem{Ren2008PoP}
Y. Ren, M. Yamada, H. Ji, S. Dorfman, S. P. Gerhardt, and R. Kulsrud, ``Experimental study of the Hall effect and electron diffusion region during magnetic reconnection in a laboratory plasma,'' Phys. Plasmas \textbf{15}, 082113 (2008), doi:10.1063/1.2936269.

\bibitem{Ji2026FLARE}
H. Ji, J. Yoo, P. Shi, \textit{et al.}, ``The FLARE facility,'' arXiv:2608.17332 (2026).

\bibitem{JiDaughton2011PoP}
H. Ji and W. Daughton, ``Phase diagram for magnetic reconnection in heliophysical, astrophysical, and laboratory plasmas,'' Phys. Plasmas \textbf{18}, 111207 (2011), doi:10.1063/1.3647505.

\bibitem{Ji2022NRP}
H. Ji, W. Daughton, J. Jara-Almonte, A. Le, A. Stanier, and J. Yoo, ``Magnetic reconnection in the era of exascale computing and multiscale experiments,'' Nat. Rev. Phys. \textbf{4}, 263 (2022), doi:10.1038/s42254-021-00419-x.

\end{thebibliography}
\end{document}